\documentclass[onecolumn,slac_one]{revtex4}
\usepackage{latexsym}
\usepackage{graphicx}
\usepackage{fancyhdr}
\renewcommand{\bar}[1]{\overline{#1}}

\newcommand{\ket}[1]{\,\left|\,{#1}\right\rangle}

\begin{document}

\title{Novel QCD Phenomena}

\author{Stanley J. Brodsky}

\affiliation{Stanford Linear Accelerator Center, Stanford University, Stanford, California 94309}

\begin{abstract} 
I discuss a number of novel topics in QCD, including the use of the AdS/CFT correspondence between Anti-de Sitter space and conformal
gauge theories to obtain an analytically tractable approximation to QCD in the regime where the QCD coupling is large and constant.  In
particular, there is an exact correspondence between the fifth-dimension coordinate $z$ of AdS space  and a specific impact variable $\zeta$
which measures the separation of the quark constituents within the hadron in ordinary space-time. This connection allows one to compute the
analytic form of the frame-independent light-front wavefunctions of mesons and baryons, the fundamental entities which encode hadron properties
and allow the computation of exclusive scattering amplitudes. I also discuss a number of novel phenomenological features of QCD. Initial- and
final-state interactions from gluon-exchange, normally neglected in the parton model, have a profound effect in QCD hard-scattering reactions,
leading to leading-twist single-spin asymmetries, diffractive deep inelastic scattering, diffractive hard hadronic reactions, the breakdown of
the Lam Tung relation in Drell-Yan reactions, and nuclear shadowing and non-universal antishadowing---leading-twist physics not incorporated in
the light-front wavefunctions of the target computed in isolation. I also discuss tests of hidden color in nuclear wavefunctions, the use of
diffraction to materialize the Fock states of a hadronic projectile and test QCD color transparency, and anomalous heavy quark effects. The
presence of direct higher-twist processes where a proton is produced in the hard subprocess can explain the large proton-to-pion ratio seen in
high centrality heavy ion collisions.
\end{abstract}

\maketitle %

\section{Introduction}
Quantum Chromodynamics is a theory with remarkably novel and interesting features. Heavy ion experiments at RHIC~\cite{Harris:2005sw} are now
discovering  unexpected new phenomena associated with the high temperature phase of QCD where its quark and gluon degrees of freedom become
manifest.  Experiments at HERMES~\cite{Airapetian:2004tw} have confirmed QCD expectations for leading-twist single-spin asymmetries which
require both the presence of quark orbital angular momentum in the proton wavefunction and  novel final-state QCD phases.  Experiments at
HERA~\cite{Derrick:1993xh} have shown that diffractive deep inelastic scattering, where the proton target remains intact, constitutes a
remarkably large percentage of the deep inelastic cross section, again showing the importance of QCD final state interactions.  The SELEX
experiment~\cite{Russ:2006me} has shown that single, and even double-charm, hadrons are produced at high $x_F$ in hadron collisions in agreement
with analyses  based on the  intrinsic charm~\cite{Brodsky:1980pb} fluctuations of the proton.  Color transparency~\cite{Brodsky:1988xz}, a key
feature of the gauge theoretic description of hadron interactions, has now been experimentally established at FermiLab~\cite{Ashery:2005vh}
using diffractive dijet production $\pi A \to {\rm jet} {\rm jet} A$.  The FermiLab experiment also provides a measurement of the valence
light-front wavefunction of the pion~\cite{Aitala:2000hb}. A similar experiment at the LHC $ p A \to Jet Jet Jet A$ at the LHC could be used to
measure the fundamental valence wavefunction of the proton~\cite{Frankfurt:2002pu}.

The LHC, in both proton-proton and heavy ion collisions, will not only open up a new high energy frontier, but it will also be a superb machine
for probing and testing QCD. The advent of new hadron physics accelerators, such as the 12 GeV electron facility at Jefferson Laboratory, the
FAIR anti-proton and heavy ion facilities at GSI, and the J-PARC hadron facility will provide many new opportunities to test QCD in its natural
domain. In addition, many novel features of QCD, such as timelike deeply virtual Compton scattering and two-photon annihilation, can be probed
at electron-positron colliders.

In this talk I will emphasize a number of aspects of QCD which seem to violate conventional wisdom:

(1) As recently noted by Collins and Qiu~\cite{Collins:2007nk}, the traditional factorization formalism of perturbative QCD for high transverse
momentum hadron production fails in detail because of initial- and final-state gluonic interactions.  These interactions produce the Sivers
effect at leading twist~\cite{Brodsky:2002cx} with different signs in semi-inclusive deep inelastic scattering and the Drell-Yan
reaction~\cite{Collins:2002kn}.  Double initial-state interactions~\cite{Boer:2002ju} also produce anomalous angular effects, including the
breakdown of the Lam-Tung relation~\cite{Lam:1980uc} in the Drell-Yan process.

(2) Hard diffractive reactions such as diffractive deep inelastic lepton scattering $e p \to e p X$  are leading-twist, Bjorken-scaling
phenomena.  In fact, as shown at HERA~\cite{Derrick:1993xh}, nearly $15\%$ of the inclusive deep inelastic cross section leaves the proton
intact. This is now understood to be due to final-state gluonic interactions of the struck quark with the proton's
spectators~\cite{Brodsky:2002ue}, contradicting models based on an intrinsic pomeron component of the proton.

(3) As emphasized by Lai, Tung, and Pumplin~\cite{Pumplin:2007wg}, there are strong indications that the structure functions used to model charm
and bottom quarks in the proton at large $x_{bj}$ have been strongly underestimated, since they ignore intrinsic heavy quark fluctuations of
hadron wavefunctions. The SELEX~\cite{Russ:2006me} discovery of $ccd$ and $ccu$ double-charm baryons at large $x_F$ reinforces other signals for
the presence of heavy quarks at large momentum fractions in hadronic wavefunctions, a rigorous feature of intrinsic heavy quark Fock states.
This has strong consequences for the production of heavy hadrons, heavy quarkonia, and even the Higgs at the LHC. Intrinsic charm and bottom
leads to substantial rates for heavy hadron production at high $x_F$~\cite{Brodsky:2006wb}, as well as anomalous nuclear effects.

(4)  The existence of dynamical higher-twist processes in which a hadron interacts directly within a hard subprocess is a rigorous prediction of
QCD.  For example, in the case of the Drell-Yan reaction $\pi p \to \ell^+ \ell^- X$  the virtual photon becomes longitudinally polarized  at
high $x_F$ , reflecting  the spin of the pion entering the QCD hard subprocess~\cite{Berger:1979du}.  In the case of high transverse momentum
proton production  the differential cross section ${d \sigma\over d^3p/E}(p p \to p p X)$ scales as $1\over p^8_T$ at fixed $x_T = 2 p_T/\sqrt
s,$~\cite{Abelev:2007ra} far from the $1/p^4_T$ to $1/p^5_T$ scaling predicted by pQCD~\cite{Brodsky:2005fz}. This suggests that the proton is
produced directly in the hard subprocess, rather than by quark or gluon fragmentation. The color transparency~\cite{Brodsky:1988xz} of the
produced proton and the resulting lack of absorption in a nuclear medium  can explain the paradoxical observation seen at RHIC that more protons
than pions are produced at high $p_T$ in high centrality heavy ion collisions.

(5) A new understanding of nuclear shadowing and antishadowing has emerged based on the presence of multi-step coherent reactions involving
leading twist diffractive reactions~\cite{Brodsky:1989qz,Brodsky:2004qa}. Thus the nuclear shadowing of structure functions is a consequence of
the lepton-nucleus collision; it is not an intrinsic property of the nuclear wavefunction. The same analysis shows that antishadowing is {\it
not universal}, but it depends in detail on the flavor of the quark or antiquark constituent~\cite{Brodsky:2004qa}.

(6) QCD predicts that a nucleus cannot be described  solely as nucleonic bound states. In the case of the deuteron, the six-quark wavefunction
has five color-singlet components, only one of which can be identified with the $p n$ state at long distances. These ``hidden color"
components~\cite{Brodsky:1983vf} play an essential role in nuclear dynamics at short distances.

(7) Spin correlations are now playing an essential role in hadron physics phenomenology, particularly in single-spin correlations which are
found to be unexpectedly strong in hadroproduction at large $x_F$ and in the double-spin correlations which measure transversity. One of the
most remarkable phenomena in hadron physics is the 4:1 ratio $R_{NN}$ of parallel to antiparallel rates seen in large-angle elastic
proton-proton scattering at $E_{cm} \simeq 5 $ GeV~\cite{Court:1986dh}. This ``exclusive transversity" is a possible signal for the existence of
$uud uud c\bar c$ resonances at the charm threshold~\cite{Brodsky:1987xw}. The absence of transverse polarization of the $J\psi$ produced at
high transverse momentum in $ p p \to J/\psi X$ is a key difficulty for heavy quark phenomenology.

(8) It is commonly believed that the renormalization scale entering the QCD coupling is an arbitrary parameter in perturbative QCD; in fact,
just as in Abelian theory, the renormalization scale is a physical quantity, representing the summation of QCD vacuum polarization contributions
to the gluon propagator in the skeleton expansion~\cite{Brodsky:1982gc,Grunberg:1991ac}.  In general, multiple renormalization scales appear in a pQCD
expression whenever multiple invariants appear in the reaction. These issues are  discussed in the next section.

These examples of unconventional wisdom highlight the need for a fundamental understanding the dynamics of hadrons in QCD at the amplitude
level. This is essential for understanding  the description of phenomena such as the quantum mechanics of hadron formation, the remarkable
effects of initial and final interactions, the origins of diffractive phenomena and single-spin asymmetries, and manifestations of higher-twist
semi-exclusive hadron subprocesses. A central tool in these analyses is the light-front wavefunctions of hadrons, the frame-independent
eigensolutions of the Heisenberg equation for QCD ~  $H^{LF}|\Psi> = M^2 |\Psi>$ quantized at fixed light-front. Given the light-front
wavefunctions $\psi_{n/H}(x_i, \vec k_{\perp i}, \lambda_i )$, one can compute a large range of exclusive and inclusive hadron observables. For
example, the valence, sea-quark and gluon distributions are defined from the squares of the LFWFS summed over all Fock states $n$. Form factors,
exclusive weak transition amplitudes~\cite{Brodsky:1998hn} such as $B\to \ell \nu \pi,$ and the generalized parton
distributions~\cite{Brodsky:2000xy} measured in deeply virtual Compton scattering are (assuming the ``handbag" approximation) overlaps of the
initial and final LFWFS with $n =n^\prime$ and $n =n^\prime+2$.

I will also discuss here a new approach~\cite{Brodsky:2006uq,Grigoryan:2007my} for determining light-front wavefunctions for QCD using the
AdS/CFT correspondence between Anti-de Sitter space and conformal gauge theories. AdS/CFT provides an analytically tractable approximation to
QCD in the regime where the QCD coupling is large and constant.  In particular, there is an exact correspondence between the fifth-dimension
coordinate $z$ of AdS space  and a specific impact variable $\zeta$ which measures the separation of the quark constituents within the hadron in
ordinary space-time. This connection allows one to compute the analytic form of the frame-independent light-front wavefunctions of mesons and
baryons, the fundamental entities which encode hadron properties and allow the computation of exclusive scattering amplitudes.

\section{Setting the Renormalization Scale in Perturbative QCD}

Precise quantitative predictions of QCD are necessary to understand the backgrounds to new beyond-the-Standard-Model phenomena at the LHC . Thus
it is important to eliminate as best as possible all uncertainties in QCD predictions, including the elimination of renormalization scale and
scheme ambiguities.

It should be emphasized that the  renormalization scale is {\it not arbitrary} in gauge theories.  For example in QED, the renormalization scale
in the usual Gell Mann-Low scheme is  exactly the photon virtuality: $\mu^2_R = k^2$. This scale sums all vacuum polarization corrections into
the dressed photon propagator of a given skeleton graph. The resulting analytic QED running coupling has dispersive cuts set correctly set at
the physical thresholds for lepton pair production $k^2= 4m^2_L$. (In ${\overline MS}$ scheme, the renormalization scales are displaced  to
$e^{-5/3} k^2$.) The renormalization scale is similarly unambiguous in QCD: the cuts due to quark loops  in the dressed gluon propagator appear
at the physical quark thresholds.  Equivalently, one can use the scheme-independent BLM
procedure~\cite{Brodsky:1982gc,Brodsky:1994eh,Rathsman:1996jk,Grunberg:1991ac} to eliminate the appearance of the $\beta$-function in the perturbative series.

Of course the {\it initial choice} of the renormalization  scale is completely arbitrary, and one can study the dependence of a perturbative
expansion on the initial scale using the usual renormalization group evolution equations.  This procedure exposes the $\beta-$dependent terms in
the PQCD expression. Eliminating the $\beta$-dependent terms then leads to a unique, physical, renormalization scale for any choice of
renormalization scheme.  In effect, one identifies the series for the corresponding conformal theory where the $\beta-$ function is zero;  the
conformal expression serves as a template~\cite{Brodsky:1999gm} for perturbative QCD expansions; the nonzero QCD $\beta$-function can then be
systematically incorporated into the scale of the running coupling~\cite{Brodsky:1994eh,Brodsky:1995tb,Brodsky:2000cr}. This leads to fixing of
the physical renormalization scale as well as commensurate scale relations which relate observables to each other without scale or scheme
ambiguity~\cite{Brodsky:1982gc}.

As an example, consider Higgs production $ p p \to H X$ calculated via $g g \to H$ fusion. The physical renormalization scale for the running
QCD couplings for  this subprocess in the pinch scheme are  the two gluon virtualities, not the Higgs mass.  The resulting values for the
renormalization scales parallel the two-photon process in QED:  $e e \to e e H$ where only vacuum polarization corrections determine the scale;
i.e., the renormalization scales are set by the photon virtualities. An interesting consequence is the prediction that the QCD coupling is
evaluated at the minimal scale of the gluon virtualities  if the Higgs is measured at $\vec p^H_T =0$.

In a physical renormalization scheme~\cite{Grunberg:1982fw}, gauge couplings are defined directly in terms of physical observables. Such
effective charges are analytic functions of the physical scales and their mass thresholds  have the correct threshold
dependence~\cite{Brodsky:1998mf,Binger:2003by} consistent with unitarity. As in QED, heavy particles contribute to physical predictions even at
energies below their threshold. This is in contrast to renormalization schemes such as $\bar{MS}$ where mass thresholds are treated as step
functions.  In the case of supersymmetric grand unification, one finds a number of qualitative differences and improvements in precision over
conventional approaches~\cite{Binger:2003by}. The analytic threshold corrections can be important in making the measured values of the gauge
couplings consistent with unification.

Relations between observables  have no scale ambiguity and are independent of the choice of the intermediate renormalization
scheme~\cite{Brodsky:1994eh};  this is the transitivity property of the renormalization group. The results, called commensurate scale relations,
are consistent~\cite{Brodsky:1992pq} with the renormalization group~\cite{Stueckelberg:1953dz} and the analytic connection  of QCD to Abelian
theory at $N_C\to 0$~\cite{Brodsky:1997jk}.  A important example is the generalized Crewther relation~\cite{Brodsky:1995tb}. One finds a
renormalization-scheme invariant relation between the coefficient function for the Bjorken sum rule for polarized deep inelastic scattering and
the $R$-ratio for the $e^+e^-$ annihilation cross section. This relation provides a generalization of the Crewther relation to non-conformally
invariant gauge theories. The derived relations allow one to calculate unambiguously without renormalization scale or scheme ambiguity the
effective charges of the polarized Bjorken and the Gross-Llewellen Smith sum rules from the experimental value for the effective charge
associated with $R$-ratio. Present data are consistent with the generalized Crewther relations, but measurements at higher precision and
energies are needed to decisively test these fundamental relations in QCD.

Recently Michael Binger and I ~\cite{Binger:2006sj} have analyzed the behavior of the thirteen nonzero form factors contributing to the
gauge-invariant three-gluon vertex at one-loop, an analysis which is important for setting the renormalization scale for heavy quark production
and other PQCD processes. Supersymmetric relations between scalar, quark, and gluon loops contributions to the triangle diagram lead to a simple
presentation of the results for a general non-Abelian gauge theories.  Only the gluon contribution to the form factors is needed since the
massless quark and scalar contributions are inferred from the homogeneous relation $F_G+4F_Q+(10-d)F_S=0$ and the sums $\Sigma_{QG}(F) \equiv
{(d-2)/ 2}F_Q + F_G$ which are given for each form factor $F$. The extension to the case of internal masses leads to the modified sum rule
$F_{MG}+4F_{MQ}+(9-d)F_{MS}=0$. The phenomenology of the three-gluon vertex is largely determined by the form factor multiplying the three-level
tensor. One can define a three-scale effective scale $Q^2_{eff}(p^2_a,p^2_b,p^2_c)$ as a function of the three external virtualities which
provides a natural extension of BLM scale setting \cite{Brodsky:1982gc} to the three-gluon vertex. Physical momentum scales thus set the scale
of the coupling. The dependence of $Q^2_{eff}$ on the physical scales has a number of surprising features.  A complicated threshold and
pseudo-threshold behavior is also observed.

\section{AdS/QCD as a First Approximant to Nonperturbative QCD}

One of the most interesting new developments in hadron physics has been the application of the AdS/CFT correspondence~\cite{Maldacena:1997re} to
nonperturbative QCD problems~\cite{Polchinski:2001tt,Janik:1999zk,Erlich:2005qh,Karch:2006pv,deTeramond:2005su}. Already AdS/CFT is giving
important insight into the  viscosity and other global properties of the hadronic system formed in heavy ion collisions~\cite{Kovtun:2004de}.

The essential ansatz for the application of AdS/CFT to hadron physics is the indication that the QCD coupling $\alpha_s(Q^2)$ becomes large and
constant in the low momentum domain $Q \le 1$ GeV/c, thus providing a window where conformal symmetry can be applied. Solutions of the QCD Dyson
Schwinger equations~\cite{vonSmekal:1997is,Zwanziger:2003cf} and phenomenological
studies~\cite{Mattingly:1993ej,Brodsky:2002nb,Baldicchi:2002qm} of QCD couplings based on physical observables such as $\tau$
decay~\cite{Brodsky:1998ua}  and the Bjorken sum rule show that the QCD $\beta$ function vanishes and  $\alpha_s(Q^2)$ become constant at small
virtuality; {\em i.e.}, effective charges develop an infrared fixed point. Recent lattice gauge theory simulations~\cite{Furui:2006py} and
nonperturbative analyses~\cite{Antonov:2007mx} have also indicated an infrared fixed point for QCD.  One can understand this
physically~\cite{Brodsky:2007pt}: in a confining theory where gluons have an effective mass or maximal wavelength, all vacuum polarization
corrections to the gluon self-energy decouple at long wavelength.  When the coupling is constant and quark masses can be ignored, the  QCD
Lagrangian becomes conformally invariant~\cite{Parisi:1972zy}, allowing the mathematically tools of conformal symmetry to be applied.

The leading power fall-off of the hard scattering amplitude as given by dimensional counting rules follows from the conformal scaling of the
underlying hard-scattering amplitude: $T_H \sim 1/Q^{n-4}$, where $n$ is the total number of fields (quarks, leptons, or gauge fields)
participating in the hard scattering~\cite{Brodsky:1974vy,Matveev:1973ra}. Thus the reaction is dominated by subprocesses and Fock states
involving the minimum number of interacting fields.  In the case of $2 \to 2$ scattering processes, this implies $ {d\sigma/ dt}(A B \to C D)
={F_{A B \to C D}(t/s)/ s^{n-2}},$ where $n = N_A + N_B + N_C +N_D$ and $n_H$ is the minimum number of constituents of $H$. The near-constancy
of the effective QCD coupling helps explain the empirical success of dimensional counting rules for the near-conformal power law fall-off of
form factors and fixed angle scaling~\cite{Brodsky:1989pv}.  For example, one sees the onset of perturbative QCD scaling behavior even for
exclusive nuclear amplitudes such as deuteron photodisintegration (Here $n = 1+ 6 + 3 + 3 = 13 .$) $s^{11}{ d\sigma/dt}(\gamma d \to p n) \sim $
constant at fixed CM angle.

The measured deuteron form factor also appears to follow the leading-twist QCD predictions at large momentum transfers in the few GeV
region~\cite{Holt:1990ze,Bochna:1998ca,Rossi:2004qm}.

Recently the Hall A collaboration at Jefferson Laboratory~\cite{Danagoulian:2007gs} has reported a significant exception to the general
empirical success of dimensional counting in fixed CM angle Compton scattering ${d\sigma\over dt}(\gamma p \to \gamma p) \sim
{F(\theta_{CM})\over s^8} $ instead of the predicted $1\over s^6$ scaling.  However, the hadron form factor $R_V(T)$, which multiplies the
$\gamma q \to \gamma q$ amplitude is found by Hall-A to scale as $1\over t^2$, in agreement with the  PQCD and AdS/CFT prediction. In addition
the timelike two-photon process $\gamma \gamma \to p \bar p$ appears to satisfy dimensional counting~\cite{Chen:2001sm,Chen:2006ug}.

The vanishing of the $\beta$ function at small momentum transfer implies that there is regime where QCD resembles a strongly-coupled theory and
mathematical techniques based on conformal invariance can be applied.  One can use the AdS/CFT correspondence between Anti-de Sitter space and
conformal gauge theories to obtain an approximation to nonperturbative QCD in the regime where the QCD coupling is large and constant; i.e., one
can use the mathematical representation of the conformal group $S0(4,2)$ in five- dimensional anti-de Sitter space to construct a holographic
representation to the theory. For example, Guy de Teramond and I~\cite{Brodsky:2006uq} have shown that the amplitude $\Phi(z)$ describing the
hadronic state in the fifth dimension of Anti-de Sitter space $\rm{AdS}_5$ can be precisely mapped to the light-front wavefunctions $\psi_{n/h}$
of hadrons in physical space-time, thus providing a description of hadrons in QCD at the amplitude level.  The light-front wavefunctions are
relativistic and frame-independent generalizations of the familiar Schr\"odinger wavefunctions of atomic physics, but they are determined at
fixed light-cone time $\tau= t +z/c$---the ``front form" advocated by Dirac---rather than at fixed ordinary time $t$. We derived this
correspondence by noticing that the mapping of $z \to \zeta$ analytically transforms the expression for the form factors in AdS/CFT to the exact
Drell-Yan-West expression in terms of light-front wavefunctions.

\begin{figure}[htb]
\centering
\includegraphics[angle=270,width=10.6cm]{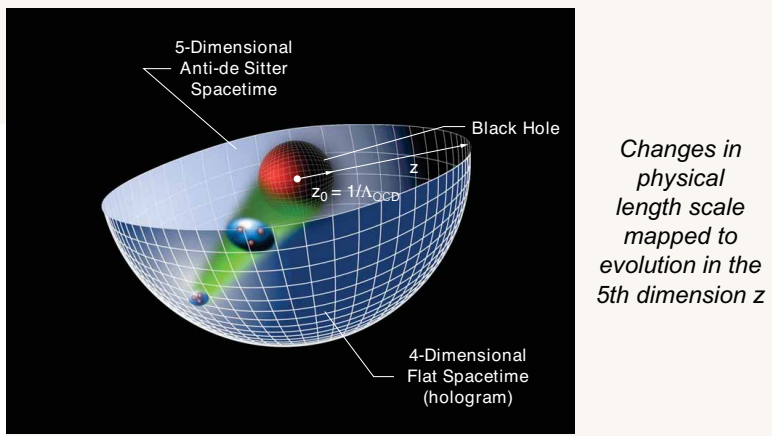}
\caption{Artist's conception of AdS/CFT.  The evolution of the
proton at different length scales is mapped into the compact $AdS_5$
dimension $z$. The black hole represents the bag-like Dirichlet
boundary condition ($\psi(z)\vert_{z= z_0={1/\Lambda_{QCD}}}=0),$
thus limiting interquark separations.}\label{fig1}
\end{figure}

Conformal symmetry can provide a systematic approximation to QCD in both its nonperturbative and perturbative domains. In the case of
nonperturbative QCD, one can use the AdS/CFT correspondence~\cite{Maldacena:1997re} between Anti-de Sitter space and conformal gauge theories to
obtain an analytically tractable approximation to QCD in the regime where the QCD coupling is large and constant.  Scale-changes in the physical
$3+1$ world can then be represented by studying dynamics in a mathematical fifth dimension with the ${\rm AdS}_5$ metric.
This is illustrated in fig. \ref{fig1}.
This connection allows one to compute the analytic form~\cite{Brodsky:2006uq,Brodsky:2007pt} of the light-front wavefunctions of mesons and
baryons.  AdS/CFT also provides a non-perturbative derivation of dimensional counting rules for the power-law fall-off of form factors and
exclusive scattering amplitudes at large momentum transfer.
The AdS/CFT approach thus allows one to construct a model of hadrons which has both confinement at large distances and the conformal scaling
properties which reproduce dimensional counting rules for hard exclusive reactions.  The fundamental equation of AdS/CFT has the appearance of a
radial Schr\"odinger Coulomb equation, but it is relativistic, covariant, and analytically tractable.

A key result from AdS/CFT  is an effective two-particle light-front radial equation for mesons~\cite{Brodsky:2006uq,Brodsky:2007pt}
\begin{equation}
\label{eq:Scheq} \left[-\frac{d^2}{d \zeta^2} + V(\zeta) \right] \phi(\zeta) = \mathcal{M}^2 \phi(\zeta),
\end{equation}
with the conformal potential $V(\zeta) = - (1-4 L^2)/4\zeta^2.$ Here  $\zeta^2 = x(1-x) \mathbf{b}_\perp^2$ where $x={k^+/ P^+}$ is the light
cone momentum fraction, and $b_\perp$ is the impact separation; i.e. the Fourier conjugate to the relative transverse momentum $k_\perp$. The
variable $\zeta$, $0 \le \zeta \le \Lambda^{-1}_{\rm QCD}$, represents the invariant separation between point-like constituents, and it is also
the holographic variable $z$ in AdS; {\em i.e.}, we can identify $\zeta = z$. The solution to (\ref{eq:Scheq}) is $\phi(z) = z^{-\frac{3}{2}}
\Phi(z) = C z^\frac{1}{2} J_L(z \mathcal{M})$. This equation reproduces the AdS/CFT solutions. The lowest stable state is determined by the
Breitenlohner-Freedman bound~\cite{Breitenlohner:1982jf}.  We can model confinement by imposing Dirichlet boundary conditions at $\phi(z =
1/\Lambda_{\rm QCD}) = 0.$ The eigenvalues are then given in terms of the roots of the Bessel functions: $\mathcal{M}_{L,k} = \beta_{L,k}
\Lambda_{\rm QCD}$.  Alternatively, one can add a confinement potential $-\kappa^2 \zeta^2$ to the effective potential
$V(\zeta)$~\cite{Karch:2006pv}.

\begin{figure}[htb]
\centering
\includegraphics[angle=270,width=0.6\textwidth]{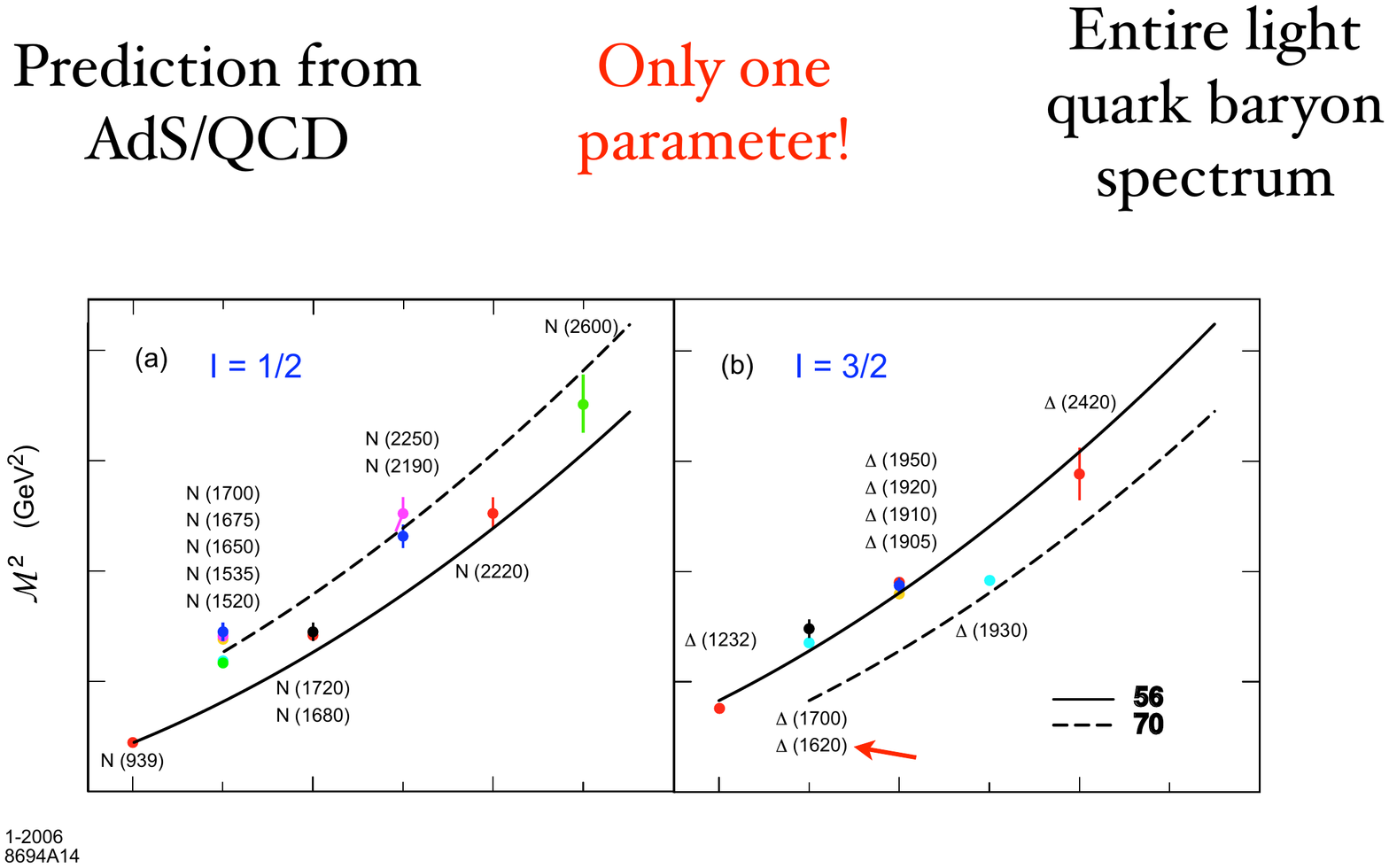}
\caption{Predictions for the masses of the orbital excitations of the $I=1/2$ and $I=3/2$ baryon states from AdS/CFT using the truncated space
model.  All four-star states listed by the Particle Data Group are shown. $\Lambda_{\rm QCD}$ =0.25GeV. The 56 trajectory corresponds to L even
P = + states, and the 70 to L odd P = - states. } \label{fig2}
\end{figure}

The eigenvalues of the effective light-front equation provide a  good description of the meson and baryon spectra for light
quarks~\cite{Brodsky:2005kc}, and its eigensolutions provide a remarkably simple but realistic model of their valence wavefunctions.  The
resulting normalized light-front wavefunctions  for the truncated space model are
\begin{equation} \widetilde \psi_{L,k}(x, \zeta) =  B_{L,k} \sqrt{x(1-x)} J_L \left(\zeta \beta_{L,k} \Lambda_{\rm QCD}\right) \theta\big(\zeta \le \Lambda^{-1}_{\rm
QCD}\big), \end{equation} where 
$B_{L,k} = \pi^{-\frac{1}{2}} {\Lambda_{\rm QCD}/ J_{1+L}(\beta_{L,k})}$. 
The results  display confinement at
large inter-quark separation and conformal symmetry at short distances, thus reproducing dimensional counting rules for hard exclusive
processes. One can also derive analogous equations for baryons composed of massless quarks using a Dirac matrix representation for the baryon
system. Predictions for the baryon spectrum are shown in fig.\ref{fig2}.

\begin{figure}[htb]
\centering
\includegraphics[angle=0,width=0.6\textwidth]{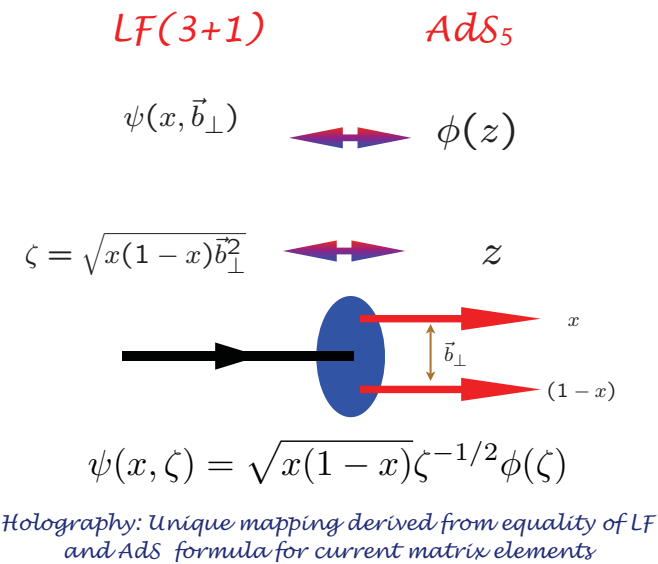}
\caption{Holographic mapping of the wavefunction $\phi(z)$ in the fifth-dimension coordinate $z$ to the light-front wavefunction as a function
of the covariant impact coordinate $\zeta = \sqrt{x(1-x)} b_\perp.$} \label{fig3}
\end{figure}

\begin{figure}[htb]
\centering
  \includegraphics[angle=0,width=0.6\textwidth]{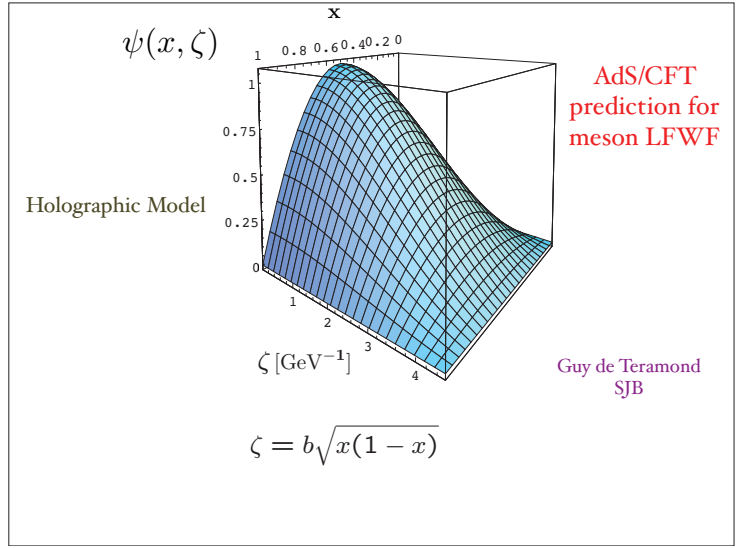}
 \caption{Illustration of the valence $q \bar q$ Fock state light-front wavefunction $\psi(x,\zeta)$ of a meson predicted by AdS/CFT.}
 \label{fig4}
 \end{figure}

\begin{figure}[htb]
\centering
\includegraphics[angle=0,width=0.6\textwidth]{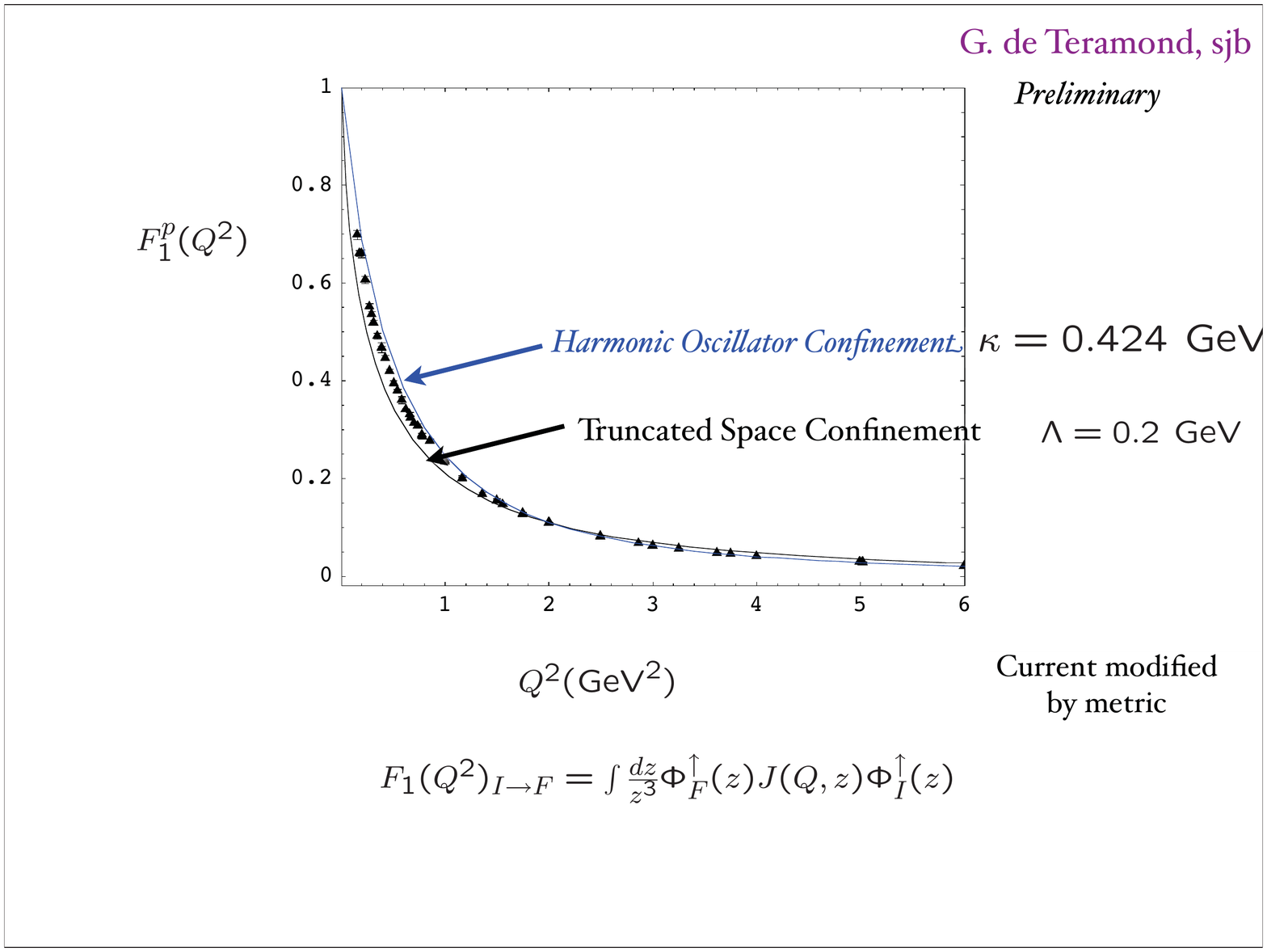}
\caption{Predictions from AdS/CFT for the space-like Dirac form factor of the proton $F_1(Q^2)$ for both the hard wall (truncated space) and
soft wall (harmonic oscillator confinement) models.  The current $J$ is modified by the metric. For example, in the soft wall model, $
J_\kappa(Q,z) = \Gamma \! \left(1+ {Q^2 \over 4 \kappa^2}\right) U \! \left({Q^2\over 4 \kappa^2}, 0 , \kappa^2 z^2\right),$ where $U(a,b,z)$ is
the confluent hypergeometric function~\cite{Brodsky:2007}.} \label{fig5}
\end{figure}

Most important, the eigensolutions of the AdS/CFT equation can be mapped to light-front equations of the hadrons in physical space-time, thus
providing an elegant description of the light hadrons at the amplitude level.  The mapping is illustrated in fig.\ref{fig3}. The meson LFWF is
illustrated in fig.\ref{fig4}.  The prediction for the proton Dirac form factor is shown in fig.\ref{fig5}.

\begin{figure}[htb]
\centering
\includegraphics[angle=0,width=0.6\textwidth]{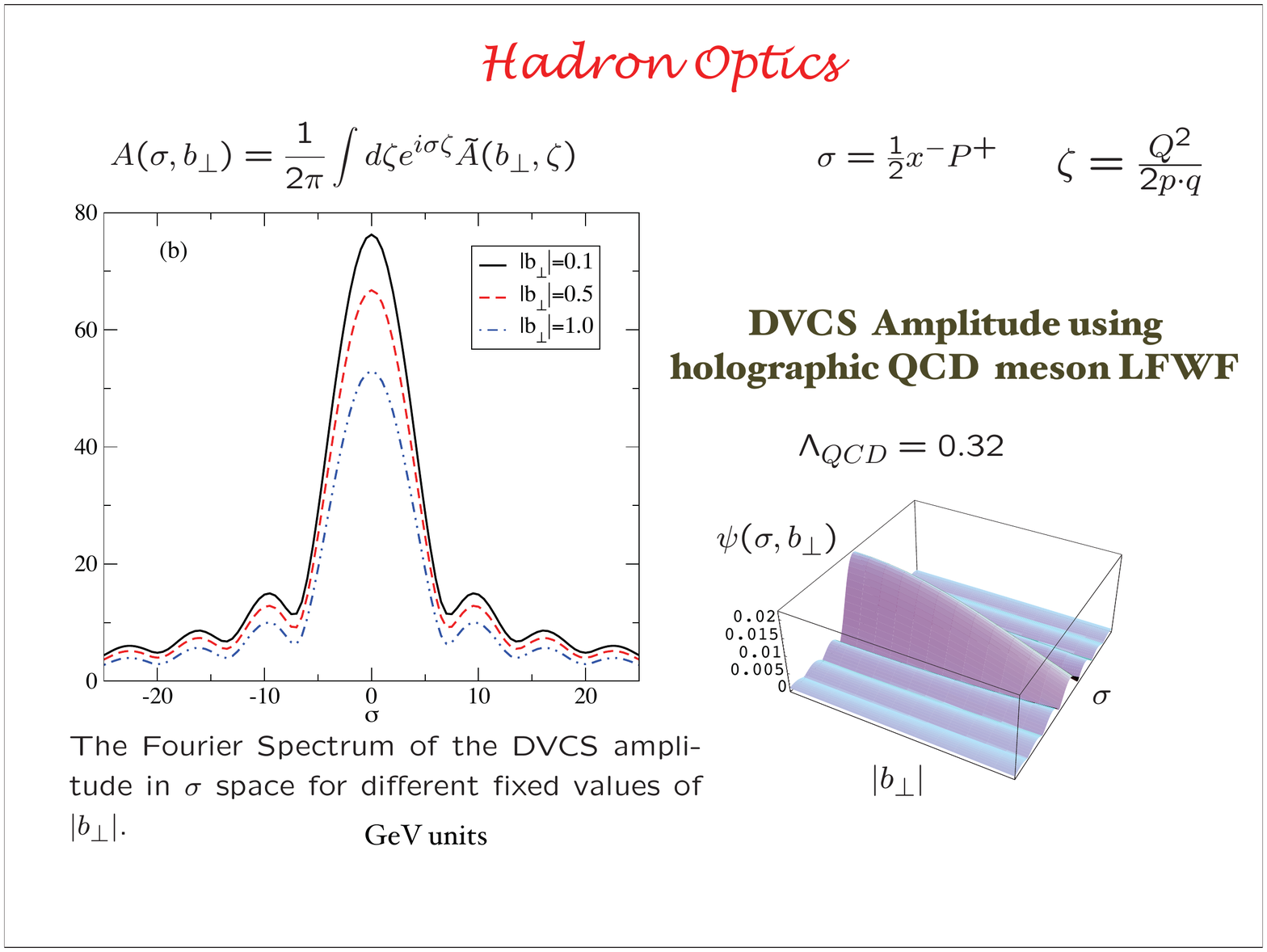}
\caption{The Fourier transform of the skewness  $\xi$ distribution of the generalized parton distribution predicted by AdS/CFT, giving
information of the hadron  in the light-front coordinate $\sigma = x^- P^+/2$ ~\cite{Brodsky:2006in}.} \label{fig6}
\end{figure}

The deeply virtual Compton amplitudes can be Fourier transformed to $b_\perp$ and $\sigma = x^-P^+/2$ space providing new insights into QCD
distributions~\cite{Burkardt:2005td,Ji:2003ak,Brodsky:2006in,Hoyer:2006xg}. The distributions in the LF direction $\sigma$ typically display
diffraction patterns arising from the interference of the initial and final state LFWFs ~\cite{Brodsky:2006in,Brodsky:2006ku}. This is
illustrated in fig.\ref{fig6}.  All of these processes  can provide a detailed test of  the AdS/CFT LFWFs predictions.

It is interesting to note that the pion distribution amplitude predicted by AdS/CFT at the hadronic scale is $\phi_\pi(x, Q _0) = {(4/ \sqrt 3
\pi)} f_\pi \sqrt{x(1-x)}$ from both the harmonic oscillator and truncated space models is quite different than the asymptotic distribution
amplitude predicted from the PQCD evolution~\cite{Lepage:1979zb} of the pion distribution amplitude: $\phi_\pi(x,Q \to \infty)= \sqrt 3  f_\pi
x(1-x) $. The broader shape of the AdS/CFT pion distribution increases the magnitude of the leading-twist perturbative QCD prediction for the
pion form factor by a factor of $16/9$ compared to the prediction based on the asymptotic form, bringing the PQCD prediction close to the
empirical pion form factor~\cite{Choi:2006ha}. Hadron form factors can be directly predicted from the overlap integrals in AdS space or
equivalently by using the Drell-Yan-West formula in physical space-time.   The form factor at high $Q^2$ receives contributions from small
$\zeta \sim {1/ Q}$, corresponding to small $\vec b_\perp$ and  $1-x$ .

The $x \to 1$ endpoint domain of structure functions is often referred to as a "soft" Feynman contribution. In fact $x \to 1$ for the struck
quark requires that all of the spectators have $x = k^+/P^+ = (k^0+ k^z)/P^+  \to 0$; this in turn requires high longitudinal momenta $k^z \to -
\infty$ for all spectators  -- unless one has both massless spectator quarks $m \equiv 0$ with zero transverse momentum $k_\perp \equiv 0$,
which is a regime of measure zero. If one uses a covariant formalism, such as the Bethe-Salpeter theory, then the virtuality of the struck quark
becomes  infinitely spacelike:  $k^2_F \sim  - {(k^2_\perp + m^2)/(1-x)}$  in the endpoint domain. Thus, actually,  $x \to 1$ corresponds to
infinite relative longitudinal momentum; it is as hard a domain in the hadron wavefunction as high transverse momentum. Note also that  at large
$x$ where the struck quark is far-off shell, DGLAP evolution is quenched~\cite{Brodsky:1979qm}, so that the fall-off of the DIS cross sections
in $Q^2$ satisfies inclusive-exclusive duality at fixed $W^2.$

The AdS/CFT approach thus provides a viable, analytic  first approximation to QCD. In principle, the model can be systematically improved, for
example by using the AdS/CFT eigensolutions as a basis for diagonalizing the full QCD Hamiltonian. An outline of the AdS/QCD program is shown in
fig.\ref{fig8}.
\begin{figure}[htb]
\centering
 \includegraphics[angle=270,width=0.6\textwidth]{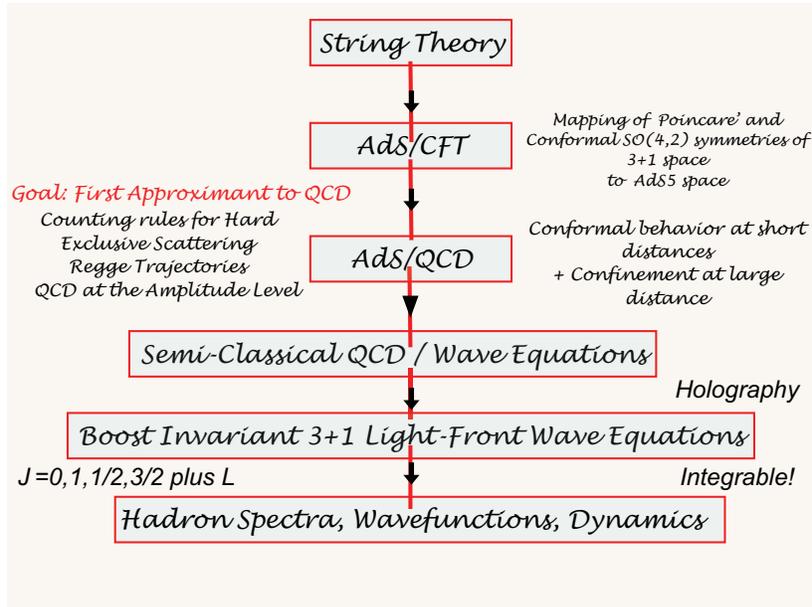}
 \caption{The logistics of AdS/CFT which leads to an analytic first approximation to QCD in its conformal window.}
 \label{fig7}
 \end{figure}
The phenomenology of the AdS/CFT model is just beginning, but it can be anticipated that it will have many applications to LHC phenomena. For
example, the model LFWFs provide a basis for understanding hadron structure functions and fragmentation functions at the amplitude level; the
same wavefunctions can describe hadron formation from the coalescence of co-moving quarks.  The spin correlations which underly single and
double spin correlations are also described by the AdS/CFT eigensolutions.  The AdS/CFT hadronic wavefunctions provides predictions for the
generalized parton distributions and weak decay amplitudes from first principles.  In addition, a prediction from AdS/CFT for the proton LFWF
would allow one to compute the higher-twist direct subprocesses such as $ u u \to p \bar d $ which could control proton production in inclusive
reactions at large transverse momenta from first principles. The amplitudes relevant to diffractive reactions could also be computed. We also
anticipate that the extension of the AdS/CFT formalism to heavy quarks will allow a great variety of heavy hadron phenomena to be analyzed from
first principles.

\section{Higher-Twist Contributions to Inclusive Reactions}
Although the contributions of higher twist processes are nominally power-law suppressed at high transverse momentum, there are some
phenomenological examples where they can play a dominant role. For example, hadrons can interact directly within a hard subprocess, leading to
higher twist contributions which can actually dominate over leading twist processes~\cite{Berger:1979du,Berger:1981fr}.  A classic example is
the reaction $\pi q \to \ell^+ \ell^- q^\prime$ which, despite its relative $1\over Q^2$ fall-off, dominates the leading twist contribution to
the Drell-Yan reaction $\pi N \to \ell^+ \ell^- X$ at high $x_F,$ producing longitudinally polarized lepton pairs.  Crossing predicts that one
also has reactions where the final-state hadron appears directly in the subprocess such as $e^+ e^- \to \pi X$ at $z=1$.

\begin{figure}[htb]
\centering
\includegraphics[angle=270,width=0.6\textwidth]{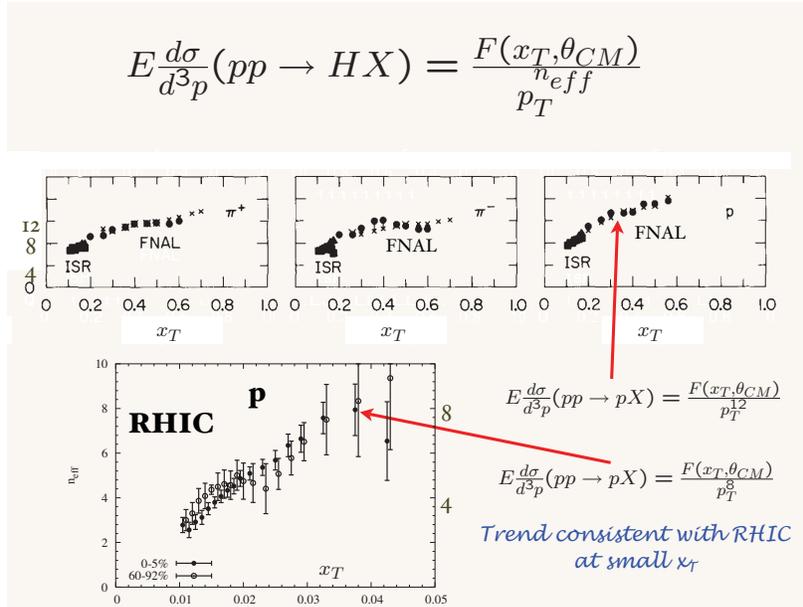}
\caption{Power-law scaling~\cite{Tannenbaum:2006ku} for hadron production at large transverse momentum from experiments at the ISR, FermiLab,
and the PHENIX collaboration at RHIC. The leading-twist prediction is $n_{eff} \simeq 4$. The $n_{eff} \sim 8$ scaling behavior observed at RHIC
for both $p A \to p X$ and $A A \to p X$ at $x_T > 0.03 $  is consistent with the dominance of a higher-twist direct process. } \label{fig8}
 \end{figure}

The fundamental test of leading-twist QCD predictions in high transverse momentum hadronic reactions is the measurement of the power-law
fall-off of the inclusive cross section ${d \sigma\over d^3p/E}(A B \to C X) ={ F(\theta_{cm}, x_T)\over p_T^{n_eff} } $ at fixed $x_T = 2
p_T/\sqrt s$ and fixed $\theta_{CM}$ where $n_{eff} \sim 4 + \delta$. Here $\delta  \le 1$ is the correction to the conformal prediction arising
from the QCD running coupling and DGLAP evolution of the input distribution and fragmentation functions~\cite{Brodsky:2005fz}. Striking
deviations from the leading-twist predictions were observed at the ISR and  Fermilab fixed-target experiments~\cite{Sivers:1975dg}. For example,
the Chicago-Princeton experiment~\cite{Antreasyan:1978cw} found $n_{eff} \simeq 12$ for $ p p \to p X$ at large, fixed  $x_T$.  A compilation of
results for the power fall-off for hard inclusive hadronic reactions is shown in fig.\ref{fig8}.

It is conventional to assume that leading-twist subprocesses dominate measurements of high $p_T$ hadron production at RHIC energies. Indeed the
data for direct photon fragmentation $ p p \to \gamma X $ is quite consistent with $n_{eff}(p p \to \gamma X) = 5 ,$  as expected from the $g q
\to \gamma q$ leading-twist subprocess. This also is likely true for pion production, at least for small $x_T.$  However, the measured fixed
$x_T$ scaling for proton production at RHIC is anomalous:  PHENIX reports $n_{eff} (p p \to p X)\simeq 8$.  A review of this data is given by
Tannenbaum~\cite{Tannenbaum:2006ku}. One can understand the anomalous scaling if a higher-twist subprocess~\cite{Brodsky:2005fz} , where the
proton is made {\it directly} within the hard reaction such as $ u u \to p \bar d$ and $(uud) u \to p u$, dominate the reaction $ p p \to p X$
at RHIC energies.  This is illustrated in fig.\ref{fig11}.
 \begin{figure}[htb]
\centering
 \includegraphics[angle=270,width=0.6\textwidth]{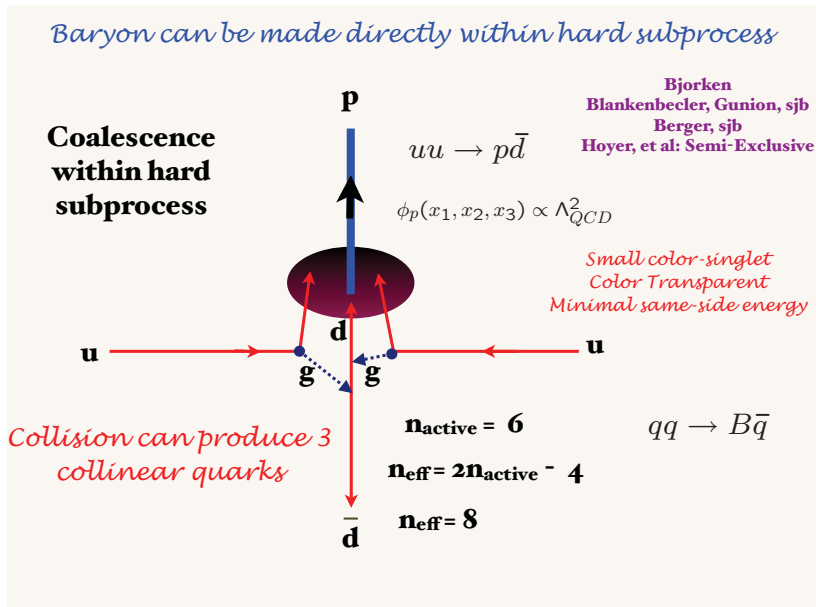}
 \caption{Representative higher-twist mechanism for direct proton production at large transverse momentum based on the subprocess $uu \to p \bar d.$
 The cross section scales as $E {d\sigma\over d^3p} = {F(x_T, \theta_{CM})\over p^{n_{eff}}_T}$ where $n_{eff} = 8$. }
 \label{fig9}
 \end{figure}
Such processes are rigorous QCD contributions. The dominance of direct subprocesses is possible since the fragmentation of gluon or quark jets
to baryons requires that the  2 to 2 subprocess occurs at much higher transverse momentum than the $p_T$ of observed proton because of the fast
falling $(1-z)^3 $ quark-to-proton fragmentation function.  Such ``direct" reactions can readily explain the fast-falling power-law  falloff
observed at fixed $x_T$ and fixed-$\theta_{cm}$ observed at the ISR, FermiLab and RHIC~\cite{Brodsky:2005fz}.  Furthermore, the protons produced
directly within the hard subprocess emerge as small-size color-transparent colored states which are not absorbed in the nuclear target. In
contrast, pions produced from jet fragmentation have the normal cross section. This provides a plausible explanation of RHIC
data~\cite{Adler:2003kg}, which shows a dramatic rise of the $p \to \pi$ ratio at high $p_T$ when one compares  peripheral  with  central (full
overlap) heavy ion collisions. This is illustrated in fig.\ref{fig10}.  The directly produced protons are not absorbed, but the pions are
diminished in the nuclear medium.
\begin{figure}[htb]
\centering
\includegraphics[angle=270,width=0.6\textwidth]{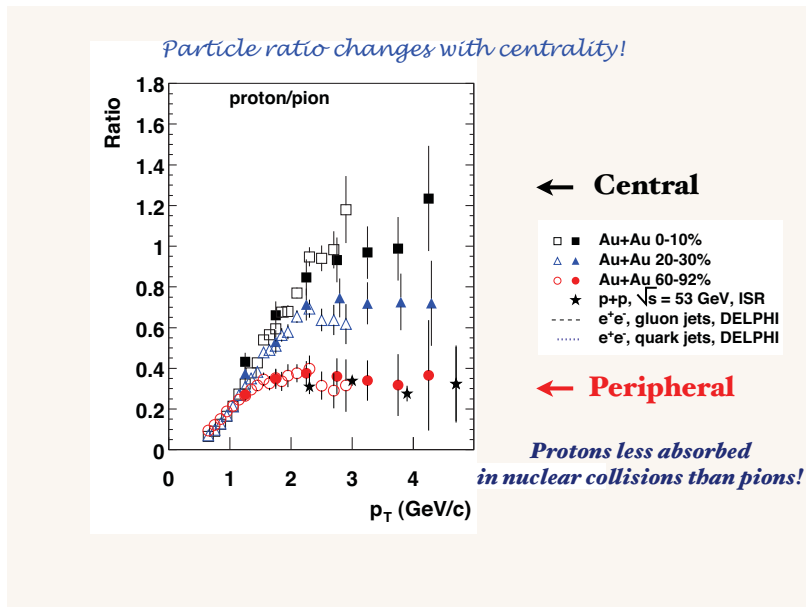}
\caption{The ratio of protons to pions produced at large $p_T$ in heavy ion collisions as a function of centrality from the PHENIX experiment at
RHIC~\cite{Adler:2003kg}.  The open and solid symbols indicate neutral versus charged pions. The rise of the $p /\pi$ ratio with $p_T$ is
consistent with the hypothesis that only the pions are absorbed in the nuclear medium.  A comparison with the measured $p/\pi$ ratio in $e^+
e^-$ and $ p p $ reactions is also shown.}
 \label{fig10}
 \end{figure}

\section{Intrinsic Heavy Quarks and the Anomalous Nuclear Dependence of Quarkonium Production}
The probability for Fock states of a light hadron such as the proton to have an extra heavy quark pair decreases as $1/m^2_Q$ in non-Abelian
gauge theory~\cite{Franz:2000ee,Brodsky:1984nx}.  The relevant matrix element is the cube of the QCD field strength $G^3_{\mu \nu}.$  This is in
contrast to abelian gauge theory where the relevant operator is $F^4_{\mu \nu}$ and the probability of intrinsic heavy leptons in QED bound
state is suppressed as $1/m^4_\ell.$  The intrinsic Fock state probability is maximized at minimal off-shellness.  It is useful to define the
transverse mass $m_{\perp i}= \sqrt{k^2_{\perp i} + m^2_i}.$ The maximum probability then occurs at $x_i = { m^i_\perp /\sum^n_{j = 1}
m^j_\perp}$; {\em i.e.}, when the constituents have minimal invariant mass and equal rapidity. Thus the heaviest constituents have the highest
momentum fractions and the highest $x_i$. Intrinsic charm thus predicts that the charm structure function has support at large $x_{bj}$ in
excess of DGLAP extrapolations~\cite{Brodsky:1980pb}; this is in agreement with the EMC measurements~\cite{Harris:1995jx}. Intrinsic charm can
also explain the $J/\psi \to \rho \pi$ puzzle~\cite{Brodsky:1997fj}. It also affects the extraction of suppressed CKM matrix elements in $B$
decays~\cite{Brodsky:2001yt}.

The dissociation of the intrinsic charm $|uud c \bar c>$ Fock state of the proton on a nucleus can produce a leading heavy quarkonium state at
high $x_F = x_c + x_{\bar c}~$ in $p A \to J/\psi X A^\prime$ since the $c$ and $\bar c$ can readily coalesce into the charmonium state.  Since
the constituents of a given intrinsic heavy-quark Fock state tend to have the same rapidity, coalescence of multiple partons from the projectile
Fock state into charmed hadrons and mesons is also favored.  For example, one can produce a leading $\Lambda_c$ at high $x_F$ and low $p_T$ from
the coalescence of the $u d c$ constituents of the projectile $|uud c \bar c>$  Fock state.  A similar coalescence mechanism was used in atomic
physics to produce relativistic antihydrogen in $\bar p A$ collisions~\cite{Munger:1993kq}. This phenomena is important not only for
understanding heavy-hadron phenomenology, but also for understanding the sources of neutrinos in astrophysics experiments~\cite{Halzen:2004bn}
and the ``long-flying" component in cosmic rays~\cite{Dremin:2005dm}.

In the case of a nuclear target, the charmonium state will be produced at small transverse momentum and high $x_F$  with a characteristic
$A^{2/3}$ nuclear dependence since the color-octet color-octet $|(uud)_{8C} (c \bar c)_{8C} >$ Fock state interacts on the front surface of the
nuclear target~\cite{Brodsky:2006wb}. This forward contribution is in addition to the $A^1$ contribution derived from the usual perturbative QCD
fusion contribution at small $x_F.$   Because of these two components, the cross section violates perturbative QCD factorization for hard
inclusive reactions~\cite{Hoyer:1990us}.  This is consistent with the observed two-component cross section for charmonium production observed by
the NA3 collaboration at CERN~\cite{Badier:1981ci} and more recent experiments~\cite{Leitch:1999ea}. The diffractive dissociation of the
intrinsic charm Fock state leads to leading charm hadron production and fast charmonium production in agreement with
measurements~\cite{Anjos:2001jr}.  Intrinsic charm can also explain the $J/\psi \to \rho \pi$ puzzle~\cite{Brodsky:1997fj}, and it affects the
extraction of suppressed CKM matrix elements in $B$ decays~\cite{Brodsky:2001yt}.

The production cross section for the double-charm $\Xi_{cc}^+$ baryon~\cite{Ocherashvili:2004hi} and the production of $J/\psi$ pairs appears to
be consistent with the diffractive dissociation and coalescence of double IC Fock states~\cite{Vogt:1995tf}. These observations provide
compelling evidence for the diffractive dissociation of complex off-shell Fock states of the projectile and contradict the traditional view that
sea quarks and gluons are always produced perturbatively via DGLAP evolution. It is also conceivable that the observations~\cite{Bari:1991ty} of
$\Lambda_b$ at high $x_F$ at the ISR in high energy $p p$  collisions could be due to the diffractive dissociation and coalescence of the
``intrinsic bottom" $|uud b \bar b>$ Fock states of the proton.

\begin{figure}[htb]
\centering
\includegraphics[angle=270,width=0.6\textwidth]{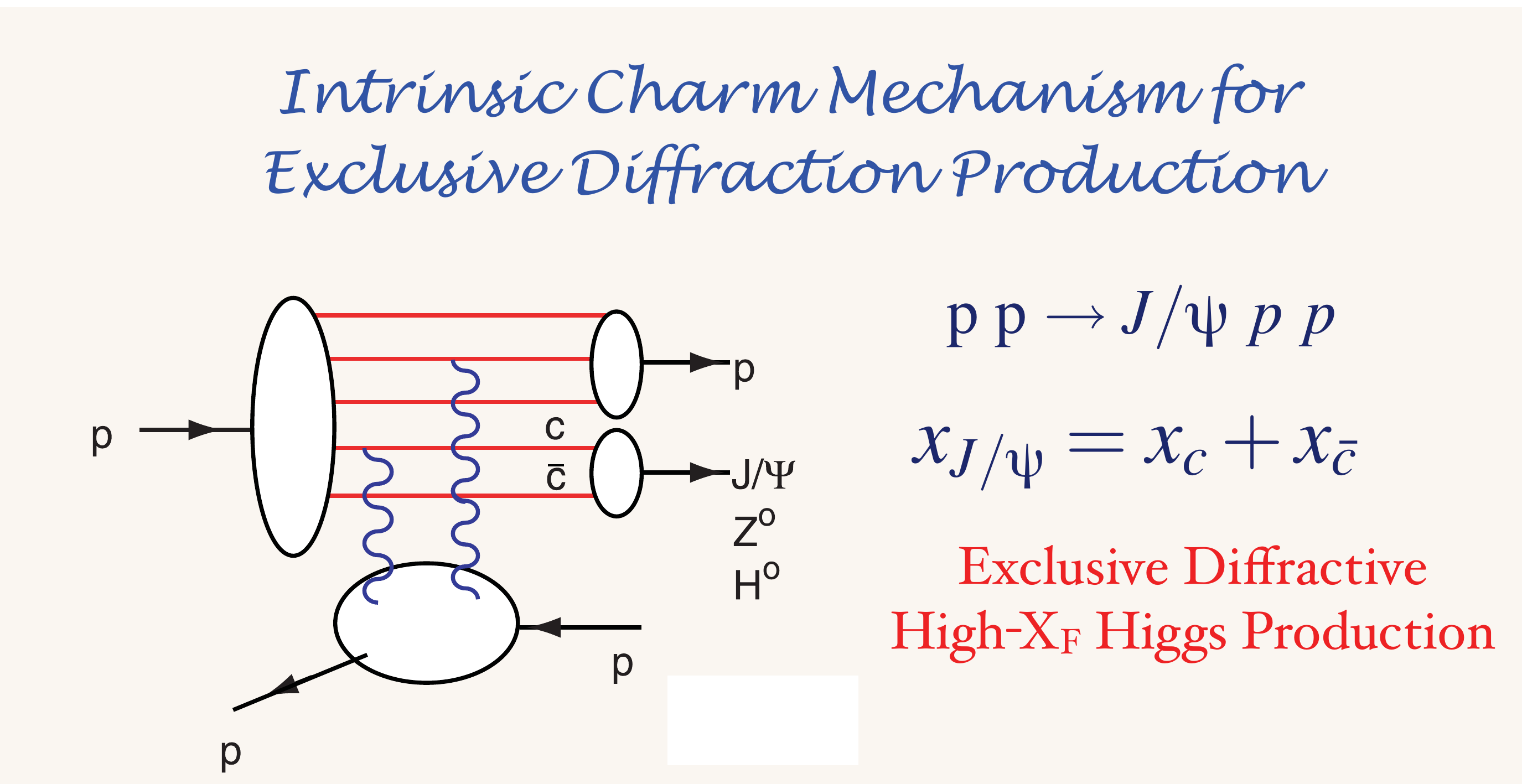}
\caption{Intrinsic charm mechanism for doubly diffractive high $x_F$ Higgs, $Z^0$ and $J/\psi $ production.  } \label{fig11}
\end{figure}

Intrinsic heavy quarks can also enhance the production probability of Higgs bosons at hadron colliders from processes such as $g c \to H c.$ It
is thus critical for new experiments (HERMES, HERA, COMPASS) to definitively establish the phenomenology of the charm structure function at
large $x_{bj}.$  Recently Kopeliovich, Schmidt, Soffer, and I ~\cite{Brodsky:2006wb} have  proposed a novel mechanism for exclusive diffractive
Higgs production $pp \to p H p $  in which the Higgs boson carries a significant fraction of the projectile proton momentum. The production
mechanism is based on the subprocess $(Q \bar Q) g \to H $ where the $Q \bar Q$ in the $|uud Q \bar Q>$ intrinsic heavy quark Fock state has up
to $80\%$ of the projectile protons momentum. This process, which is illustrated in fig.\ref{fig11},  will provide a clear experimental signal
for Higgs production due to the small background in this kinematic region.

\section{Hidden Color}
In traditional nuclear physics, the deuteron is a bound state of a proton and a neutron where the binding force arise from the exchange of a
pion and other mesonic states. However, QCD provides a new perspective:~\cite{Brodsky:1976rz,Matveev:1977xt}  six quarks in the fundamental
$3_C$ representation of $SU(3)$ color can combine into five different color-singlet combinations, only one of which corresponds to a proton and
neutron.  In fact, if the deuteron wavefunction is a proton-neutron bound state at large distances, then as their separation becomes smaller,
the QCD evolution resulting from colored gluon exchange introduce four other ``hidden color" states into the deuteron
wavefunction~\cite{Brodsky:1983vf}. The normalization of the deuteron form factor observed at large $Q^2$~\cite{Arnold:1975dd}, as well as the
presence of two mass scales in the scaling behavior of the reduced deuteron form factor~\cite{Brodsky:1976rz}, thus suggest sizable hidden-color
Fock state contributions such as $\ket{(uud)_{8_C} (ddu)_{8_C}}$ with probability  of order $15\%$ in the deuteron
wavefunction~\cite{Farrar:1991qi}.

The hidden color states of the deuteron can be materialized at the hadron level as \\  $\Delta^{++}(uuu)\Delta^{-}(ddd)$ and other novel quantum
fluctuations of the deuteron. These dual hadronic components become more and more important as one probes the deuteron at short distances, such
as in exclusive reactions at large momentum transfer.  For example, the ratio \\ ${{d \sigma/ dt}(\gamma d \to \Delta^{++}
\Delta^{-})/{d\sigma/dt}(\gamma d\to n p) }$ should increase dramatically to  a fixed ratio $2::5$ with increasing transverse momentum $p_T.$
Similarly the Coulomb dissociation of the deuteron into various exclusive channels $e d \to e^\prime + p n, p p \pi^-, \Delta \Delta, \cdots$
should have a changing composition as the final-state hadrons are probed at high transverse momentum, reflecting the onset of hidden color
degrees of freedom.

Recently the CLEO collaboration~\cite{Asner:2006pw} has measured the branching ratios of \\ $\Upsilon(nS) \to \ {\textrm{aantideuteron}}\ X$. This
reaction should be sensitive to the hidden-color structure of the anti-deuteron wavefunction since the $\Upsilon \to b \bar b \to ggg \to qqqqqq
\bar q \bar q \bar q \bar q \bar q \bar q$ originates from a system of small compact size and leads to multi-quark states with diverse colors.
It is crucial to also have data on $ \Upsilon \to \bar p \bar n X$ where the anti-nucleons emerge at minimal invariant mass.  The conventional
nuclear physics expectation can then be computed from the convolution of this distribution with the square of the two nucleon ``body"  LFWF of
the deuteron:
\begin{equation}\int d^2 k_\perp \int^1_0 dx  | \psi^{\bar d}_{\bar n  \bar p}(x,k_\perp)|^2  \times {d\sigma\over d^3 p_{\bar
n}/E_{\bar n} ~ d^3 p_{\bar p} /E_{\bar p}}(\Upsilon \to \bar n \bar p X)
\end{equation}

\section{Diffractive Deep Inelastic Scattering}
A remarkable feature of deep inelastic lepton-proton scattering at HERA is that approximately 10\% events are
diffractive~\cite{Adloff:1997sc,Breitweg:1998gc}: the target proton remains intact, and there is a large rapidity gap between the proton and the
other hadrons in the final state.  These diffractive deep inelastic scattering (DDIS) events can be understood most simply from the perspective
of the color-dipole model: the $q \bar q$ Fock state of the high-energy virtual photon diffractively dissociates into a diffractive dijet
system.  The exchange of multiple gluons between  the color dipole of the $q \bar q$ and the quarks of the target proton neutralizes the color
separation and leads to the diffractive final state.  The same multiple gluon exchange also controls diffractive vector meson electroproduction
at large photon virtuality \cite{Brodsky:1994kf}.  This observation presents a paradox: if one chooses the conventional parton model frame where
the photon light-front momentum is negative $q+ = q^0 + q^z  < 0$, the virtual photon interacts with a quark constituent with light-cone
momentum fraction $x = {k^+/p^+} = x_{bj}.$  Furthermore, the gauge link associated with the struck quark (the Wilson line) becomes unity in
light-cone gauge $A^+=0$. Thus the struck ``current" quark apparently experiences no final-state interactions. Since the light-front
wavefunctions $\psi_n(x_i,k_{\perp i})$ of a stable hadron are real, it appears impossible to generate the required imaginary phase associated
with pomeron exchange, let alone large rapidity gaps.

This paradox was resolved by Hoyer, Marchal,  Peigne, Sannino and myself \cite{Brodsky:2002ue}.  Consider the case where the virtual photon
interacts with a strange quark---the $s \bar s$ pair is assumed to be produced in the target by gluon splitting.  In the case of Feynman gauge,
the struck $s$ quark continues to interact in the final state via gluon exchange as described by the Wilson line. The final-state interactions
occur at a light-cone time $\Delta\tau \simeq 1/\nu$ shortly after the virtual photon interacts with the struck quark. When one integrates over
the nearly-on-shell intermediate state, the amplitude acquires an imaginary part. Thus the rescattering of the quark produces a separated
color-singlet $s \bar s$ and an imaginary phase. In the case of the light-cone gauge $A^+ = \eta \cdot A =0$, one must also consider the
final-state interactions of the (unstruck) $\bar s$ quark. The gluon propagator in light-cone gauge $d_{LC}^{\mu\nu}(k) = (i/k^2+ i
\epsilon)\left[-g^{\mu\nu}+\left(\eta^\mu k^\nu+ k^\mu\eta^\nu / \eta\cdot k\right)\right] $ is singular at $k^+ = \eta\cdot k = 0.$ The
momentum of the exchanged gluon $k^+$ is of ${ \cal O}{(1/\nu)}$; thus rescattering contributes at leading twist even in light-cone gauge. The
net result is  gauge invariant and is identical to the color dipole model calculation. The calculation of the rescattering effects on DIS in
Feynman and light-cone gauge through three loops is given in detail for an Abelian model in reference~\cite{Brodsky:2002ue}.  The result shows
that the rescattering corrections reduce the magnitude of the DIS cross section in analogy to nuclear shadowing.

A new understanding of the role of final-state interactions in deep inelastic scattering has thus emerged. The multiple scattering of the struck
parton via instantaneous interactions in the target generates dominantly imaginary diffractive amplitudes, giving rise to an effective ``hard
pomeron" exchange.  The presence of a rapidity gap between the target and diffractive system requires that the target remnant emerges in a
color-singlet state; this is made possible in any gauge by the soft rescattering.  The resulting diffractive contributions leave the target
intact  and do not resolve its quark structure; thus there are contributions to the DIS structure functions which cannot be interpreted as
parton probabilities~\cite{Brodsky:2002ue}; the leading-twist contribution to DIS  from rescattering of a quark in the target is a coherent
effect which is not included in the light-front wave functions computed in isolation. One can augment the light-front wave functions with a
gauge link corresponding to an external field created by the virtual photon $q \bar q$ pair current~\cite{Belitsky:2002sm,Collins:2004nx}.  Such
a gauge link is process dependent~\cite{Collins:2002kn}, so the resulting augmented LFWFs are not universal
\cite{Brodsky:2002ue,Belitsky:2002sm,Collins:2003fm}.   We also note that the shadowing of nuclear structure functions is due to the destructive
interference between multi-nucleon amplitudes involving diffractive DIS and on-shell intermediate states with a complex phase. In contrast, the
wave function of a stable target is strictly real since it does not have on-energy-shell intermediate state configurations.  The physics of
rescattering and shadowing is thus not included in the nuclear light-front wave functions, and a probabilistic interpretation of the nuclear DIS
cross section is precluded.

Rikard Enberg, Paul Hoyer, Gunnar Ingelman and I~\cite{Brodsky:2004hi} have shown that the quark structure function of the effective hard
pomeron has the same form as the quark contribution of the gluon structure function. The hard pomeron is not an intrinsic part of the proton;
rather it must be considered as a dynamical effect of the lepton-proton interaction. Our QCD-based picture also applies to diffraction in
hadron-initiated processes. The rescattering is different in virtual photon- and hadron-induced processes due to the different color
environment, which accounts for the  observed non-universality of diffractive parton distributions. This framework also provides a theoretical
basis for the phenomenologically successful Soft Color Interaction (SCI) model~\cite{Edin:1995gi} which includes rescattering effects and thus
generates a variety of final states with rapidity gaps.

\section{ Single-Spin Asymmetries from Final-State
Interactions}

\begin{figure}[htb]
\centering
\includegraphics[angle=270,width=0.6\textwidth]{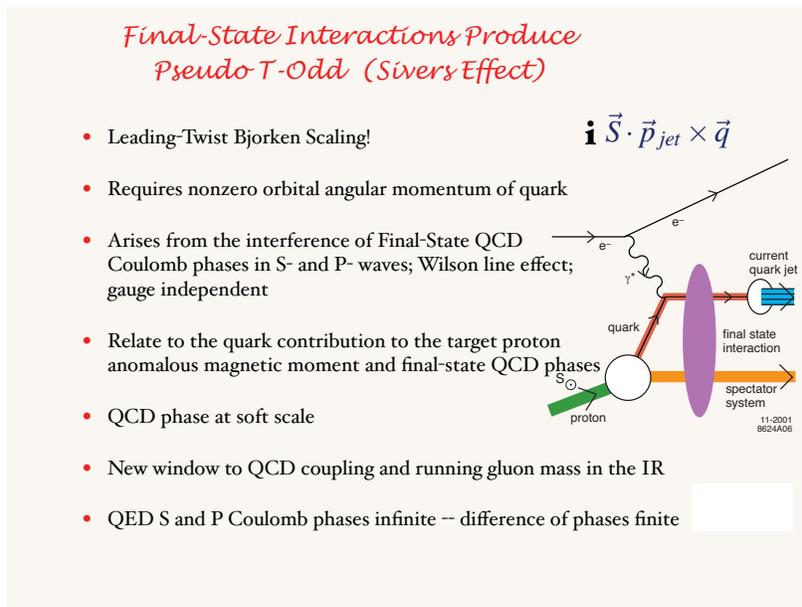}
\caption{Final-state interactions in QCD and the physics of the leading-twist Sivers single-spin asymmetry in semi-inclusive deep inelastic
lepton-proton scattering.} \label{fig12}
\end{figure}

Among the most interesting polarization effects are single-spin azimuthal asymmetries  in semi-inclusive deep inelastic scattering, representing
the correlation of the spin of the proton target and the virtual photon to hadron production plane: $\vec S_p \cdot \vec q \times \vec p_H$.
Such asymmetries are time-reversal odd, but they can arise in QCD through phase differences in different spin amplitudes. In fact, final-state
interactions from gluon exchange between the outgoing quarks and the target spectator system lead to single-spin asymmetries in semi-inclusive
deep inelastic lepton-proton scattering  which  are not power-law suppressed at large photon virtuality $Q^2$ at fixed
$x_{bj}$~\cite{Brodsky:2002cx}.  In contrast to the SSAs arising from transversity and the Collins fragmentation function, the fragmentation of
the quark into hadrons is not necessary; one predicts a correlation with the production plane of the quark jet itself. Physically, the
final-state interaction phase arises as the infrared-finite difference of QCD Coulomb phases for hadron wave functions with differing orbital
angular momentum.  See fig.\ref{fig12}. The same proton matrix element which determines the spin-orbit correlation $\vec S \cdot \vec L$ also
produces the anomalous magnetic moment of the proton, the Pauli form factor, and the generalized parton distribution $E$ which is measured in
deeply virtual Compton scattering. Thus the contribution of each quark current to the SSA is proportional to the contribution $\kappa_{q/p}$ of
that quark to the proton target's anomalous magnetic moment $\kappa_p = \sum_q e_q \kappa_{q/p}$ ~\cite{Brodsky:2002cx,Burkardt:2004vm}.  The
HERMES collaboration has recently measured the SSA in pion electroproduction using transverse target polarization~\cite{Airapetian:2004tw}. The
Sivers and Collins effects can be separated using planar correlations; both contributions are observed to contribute, with values not in
disagreement with theory expectations ~\cite{Airapetian:2004tw,Avakian:2004qt}. A related analysis also predicts that the initial-state
interactions from gluon exchange between the incoming quark and the target spectator system lead to leading-twist single-spin asymmetries in the
Drell-Yan process $H_1 H_2^\updownarrow \to \ell^+ \ell^- X$ ~\cite{Collins:2002kn,Brodsky:2002rv}.  The SSA in the Drell-Yan process is the
same as that obtained in SIDIS, with the appropriate identification of variables, but with the opposite sign. There is no Sivers effect in
charged-current reactions since the $W$ only couples to left-handed quarks~\cite{Brodsky:2002pr}.

If both the quark and antiquark in the initial state of the Drell-Yan subprocess $q \bar q \to  \mu^+ \mu^-$ interact with the spectators of the
other incident hadron, one finds a breakdown of the Lam-Tung relation, which was formerly believed to be a general prediction of leading-twist
QCD. These double initial-state interactions also lead to a $\cos 2 \phi$ planar correlation in unpolarized Drell-Yan
reactions~\cite{Boer:2002ju}. More generally one must consider subprocesses involving initial-state gluons such as $n g q \bar q \to \ell \bar
\ell$  in addition to subprocesses with extra final-state gluons.

The final-state interaction mechanism provides an appealing physical explanation within QCD of single-spin asymmetries.  Remarkably, the same
matrix element which determines the spin-orbit correlation $\vec S \cdot \vec L$ also produces the anomalous magnetic moment of the proton, the
Pauli form factor, and the generalized parton distribution $E$ which is measured in deeply virtual Compton scattering.  Physically, the
final-state interaction phase arises as the infrared-finite difference of QCD Coulomb phases for hadron wave functions with differing orbital
angular momentum.  An elegant discussion of the Sivers effect including its sign has been given by Burkardt~\cite{Burkardt:2004vm}. As shown
recently by Gardner and myself~\cite{Brodsky:2006ha}, one can also use the Sivers effect to study the orbital angular momentum of  gluons by
tagging a gluon jet in semi-inclusive DIS. In this case, the final-state interactions are enhanced by the large color charge of the gluons.

The final-state interaction effects can also be identified with the gauge link which is present in the gauge-invariant definition of parton
distributions~\cite{Collins:2004nx}.  Even when the light-cone gauge is chosen, a transverse gauge link is required.  Thus in any gauge the
parton amplitudes need to be augmented by an additional eikonal factor incorporating the final-state interaction and its
phase~\cite{Ji:2002aa,Belitsky:2002sm}. The net effect is that it is possible to define transverse momentum dependent parton distribution
functions which contain the effect of the QCD final-state interactions.

\section{Diffraction Dissociation as a Tool to Resolve Hadron
Substructure and Test Color Transparency}

Diffractive multi-jet production in heavy nuclei provides a novel way to resolve the shape of light-front Fock state wave functions and test
color transparency~\cite{Brodsky:1988xz}.  For example, consider the reaction~\cite{Bertsch:1981py,Frankfurt:1999tq}.   $\pi A \rightarrow {\rm
Jet}_1 + {\rm Jet}_2 + A^\prime$ at high energy where the nucleus $A^\prime$ is left intact in its ground state. The transverse momenta of the
jets balance so that $ \vec k_{\perp i} + \vec k_{\perp 2} = \vec q_\perp < {R^{-1}}_A \ $.  Because of color transparency, the valence wave
function of the pion with small impact separation will penetrate the nucleus with minimal interactions, diffracting into jet
pairs~\cite{Bertsch:1981py}.  The $x_1=x$, $x_2=1-x$ dependence of the dijet distributions will thus reflect the shape of the pion valence
light-cone wave function in $x$; similarly, the $\vec k_{\perp 1}- \vec k_{\perp 2}$ relative transverse momenta of the jets gives key
information on the second transverse momentum derivative of the underlying shape of the valence pion
wavefunction~\cite{Frankfurt:1999tq,Nikolaev:2000sh}. The diffractive nuclear amplitude extrapolated to $t = 0$ should be linear in nuclear
number $A$ if color transparency is correct.  The integrated diffractive rate will then scale as $A^2/R^2_A \sim A^{4/3}$. This is in fact what
has been observed by the E791 collaboration at FermiLab for 500 GeV incident pions on nuclear targets~\cite{Aitala:2000hc}.  The measured
momentum fraction distribution of the jets with high transverse momentum is found to be approximately consistent with the shape of the pion
asymptotic distribution amplitude, $\phi^{\rm asympt}_\pi (x) = \sqrt 3 f_\pi x(1-x)$~\cite{Aitala:2000hb}; however, there is an indication from
the data that the distribution is broader at lower transverse momentum, consistent with the AdS/CFT prediction.

Color transparency, as evidenced by the Fermilab measurements of diffractive dijet production, implies that a pion can interact coherently
throughout a nucleus with minimal absorption, in dramatic contrast to traditional Glauber theory based on a fixed $\sigma_{\pi n}$ cross
section.  Color transparency gives direct validation of the gauge interactions of QCD. Color transparency has also been observed in diffractive
electroproduction of $\rho$ mesons~\cite{Borisov:2002rd} and in quasi-elastic $p A \to p p (A-1)$ scattering~\cite{Aclander:2004zm} where only
the small size fluctuations of the hadron wavefunction enters the hard exclusive scattering amplitude.  In the latter case an anomaly occurs at
$\sqrt s \simeq 5 $ GeV, most likely signaling a resonance effect at the charm threshold~\cite{Brodsky:1987xw}.

\section{Shadowing and Antishadowing of Nuclear Structure Functions}

One of the novel features of QCD involving nuclei is the {\it antishadowing} of the nuclear structure functions which is observed in deep
inelastic lepton scattering and other hard processes. Empirically, one finds $R_A(x,Q^2) \equiv  \left(F_{2A}(x,Q^2)/ (A/2) F_{d}(x,Q^2)\right)
> 1 $ in the domain $0.1 < x < 0.2$; {\em i.e.}, the measured nuclear structure function (referenced to the deuteron) is larger than than the
scattering on a set of $A$ independent nucleons.

The shadowing of the nuclear structure functions: $R_A(x,Q^2) < 1 $ at small $x < 0.1 $ can be readily understood in terms of the Gribov-Glauber
theory.  Consider a two-step process in the nuclear target rest frame. The incoming $q \bar q$ dipole first interacts diffractively $\gamma^*
N_1 \to (q \bar q) N_1$ on nucleon $N_1$ leaving it intact.  This is the leading-twist diffractive deep inelastic scattering  (DDIS) process
which has been measured at HERA to constitute approximately 10\% of the DIS cross section at high energies.  The $q \bar q$ state then interacts
inelastically on a downstream nucleon $N_2:$ $(q \bar q) N_2 \to X$. The phase of the pomeron-dominated DDIS amplitude is close to imaginary,
and the Glauber cut provides another phase $i$, so that the two-step process has opposite  phase and  destructively interferes with the one-step
DIS process $\gamma* N_2 \to X$ where $N_1$ acts as an unscattered spectator. The one-step and-two step amplitudes can coherently interfere as
long as the momentum transfer to the nucleon $N_1$ is sufficiently small that it remains in the nuclear target;  {\em i.e.}, the Ioffe
length~\cite{Ioffe:1969kf} $L_I = { 2 M \nu/ Q^2} $ is large compared to the inter-nucleon separation. In effect, the flux reaching the interior
nucleons is diminished, thus reducing the number of effective nucleons and $R_A(x,Q^2) < 1.$

There are also leading-twist diffractive contributions $\gamma^* N_1 \to (q \bar q) N_1$  arising from Reggeon exchanges in the
$t$-channel~\cite{Brodsky:1989qz}.  For example, isospin--non-singlet $C=+$ Reggeons contribute to the difference of proton and neutron
structure functions, giving the characteristic Kuti-Weisskopf $F_{2p} - F_{2n} \sim x^{1-\alpha_R(0)} \sim x^{0.5}$ behavior at small $x$. The
$x$ dependence of the structure functions reflects the Regge behavior $\nu^{\alpha_R(0)} $ of the virtual Compton amplitude at fixed $Q^2$ and
$t=0.$ The phase of the diffractive amplitude is determined by analyticity and crossing to be proportional to $-1+ i$ for $\alpha_R=0.5,$ which
together with the phase from the Glauber cut, leads to {\it constructive} interference of the diffractive and nondiffractive multi-step nuclear
amplitudes. Furthermore, because of its $x$ dependence, the nuclear structure function is enhanced precisely in the domain $0.1 < x <0.2$ where
antishadowing is empirically observed.  The strength of the Reggeon amplitudes is fixed by the fits to the nucleon structure functions, so there
is little model dependence.

As noted above, the Bjorken-scaling diffractive contribution to DIS arises from the rescattering of the struck quark after it is struck  (in the
parton model frame $q^+ \le 0$), an effect induced by the Wilson line connecting the currents. Thus one cannot attribute DDIS to the physics of
the target nucleon computed in isolation~\cite{Brodsky:2002ue}.  Similarly, since shadowing and antishadowing arise from the physics of
diffraction, we cannot attribute these phenomena to the structure of the nucleus itself: shadowing and antishadowing arise because of the
$\gamma^* A$ collision and the history of the $q \bar q$ dipole as it propagates through the nucleus.

Ivan Schmidt, Jian-Jun Yang, and I~\cite{Brodsky:2004qa} have extended the Glauber analysis to the shadowing and antishadowing of all of the
electroweak structure functions. Quarks of different flavors  will couple to different Reggeons; this leads to the remarkable prediction that
nuclear antishadowing is not universal; it depends on the quantum numbers of the struck quark. This picture implies substantially different
antishadowing for charged and neutral current reactions, thus affecting the extraction of the weak-mixing angle $\theta_W$.  We find that part
of the anomalous NuTeV result~\cite{Zeller:2001hh} for $\theta_W$ could be due to the non-universality of nuclear antishadowing for charged and
neutral currents. Detailed measurements of the nuclear dependence of individual quark structure functions are thus needed to establish the
distinctive phenomenology of shadowing and antishadowing and to make the NuTeV results definitive. Schmidt, Yang, and I have also identified
contributions to the nuclear multi-step reactions which arise from odderon exchange and hidden color degrees of freedom in the nuclear
wavefunction. There are other ways in which this new view of antishadowing can be tested;  antishadowing can also depend on the target and beam
polarization.

\begin{acknowledgments}
Presented at the Workshop, ``High-pT physics at the LHC", March 23-27, 2007 in Jyvaskyla, Finland. I thank the organizers for their invitation,
especially Professor Jan Rak, to speak at this workshop.  I also thank my collaborators for many helpful discussions. This work was supported in
part by the Department of Energy, contract No. DE-AC02-76SF00515.
\end{acknowledgments}


\begin{thebibliography}{99}


\bibitem{Harris:2005sw}
  J.~W.~Harris,
  Czech.\ J.\ Phys.\  {\bf 55}, B297 (2005).

\bibitem{Airapetian:2004tw}
  A.~Airapetian {\it et al.}  [HERMES Collaboration],
  Phys.\ Rev.\ Lett.\  {\bf 94}, 012002 (2005)
  [arXiv:hep-ex/0408013].

\bibitem{Derrick:1993xh}
  M.~Derrick {\it et al.}  [ZEUS Collaboration],
  Phys.\ Lett.\  B {\bf 315}, 481 (1993).

\bibitem{Russ:2006me}
  J.~S.~Russ,
  Int.\ J.\ Mod.\ Phys.\  A {\bf 21}, 5482 (2006).

\bibitem{Brodsky:1980pb}
  S.~J.~Brodsky, P.~Hoyer, C.~Peterson and N.~Sakai,
  Phys.\ Lett.\  B {\bf 93}, 451 (1980).

\bibitem{Brodsky:1988xz}
  S.~J.~Brodsky and A.~H.~Mueller,
  Phys.\ Lett.\  B {\bf 206}, 685 (1988).

\bibitem{Ashery:2005vh}
  D.~Ashery,
  ``Measurement of light cone wave functions by diffractive
  dissociation,''
{\it Talk given at Particles and Nuclei International Conference (PANIC      05), Santa Fe, New Mexico, 24-28 Oct 2005}

\bibitem{Aitala:2000hb}
  E.~M.~Aitala {\it et al.}  [E791 Collaboration],
  Phys.\ Rev.\ Lett.\  {\bf 86}, 4768 (2001)
  [arXiv:hep-ex/0010043].

\bibitem{Frankfurt:2002pu}
  L.~Frankfurt, G.~A.~Miller and M.~Strikman,
  ``Nuclear transparency and light cone pion wave function in the coherent pion
  nucleon and pion nucleus production of two jets at high relative momenta,''
{\it Prepared for Exclusive Processes at High Momentum Transfer,
Newport News, Virginia, 15-18 May 2002}

\bibitem{Collins:2007nk}
  J.~Collins and J.~W.~Qiu,
  arXiv:0705.2141 [hep-ph].

\bibitem{Brodsky:2002cx}
  S.~J.~Brodsky, D.~S.~Hwang and I.~Schmidt,
  Int.\ J.\ Mod.\ Phys.\  A {\bf 18}, 1327 (2003)
  [Phys.\ Lett.\  B {\bf 530}, 99 (2002)]
  [arXiv:hep-ph/0201296].

\bibitem{Collins:2002kn}
  J.~C.~Collins,
  Phys.\ Lett.\  B {\bf 536}, 43 (2002)
  [arXiv:hep-ph/0204004].

\bibitem{Boer:2002ju}
  D.~Boer, S.~J.~Brodsky and D.~S.~Hwang,
  Phys.\ Rev.\  D {\bf 67}, 054003 (2003)
  [arXiv:hep-ph/0211110].

\bibitem{Lam:1980uc}
  C.~S.~Lam and W.~K.~Tung,
  Phys.\ Rev.\  D {\bf 21}, 2712 (1980).

\bibitem{Brodsky:2002ue}
  S.~J.~Brodsky, P.~Hoyer, N.~Marchal, S.~Peigne and F.~Sannino,
  Phys.\ Rev.\  D {\bf 65}, 114025 (2002)
  [arXiv:hep-ph/0104291].

\bibitem{Pumplin:2007wg}
  J.~Pumplin, H.~L.~Lai and W.~K.~Tung,
  Phys.\ Rev.\  D {\bf 75}, 054029 (2007)
  [arXiv:hep-ph/0701220].

\bibitem{Brodsky:2006wb}
  S.~J.~Brodsky, B.~Kopeliovich, I.~Schmidt and J.~Soffer,
  Phys.\ Rev.\  D {\bf 73}, 113005 (2006)
  [arXiv:hep-ph/0603238].

\bibitem{Berger:1979du}
  E.~L.~Berger and S.~J.~Brodsky,
  Phys.\ Rev.\ Lett.\  {\bf 42}, 940 (1979).

\bibitem{Abelev:2007ra}
  B.~I.~Abelev {\it et al.}  [STAR Collaboration],
  arXiv:nucl-ex/0703040.


\bibitem{Brodsky:2005fz}
  S.~J.~Brodsky and M.~Rijssenbeek,
  arXiv:hep-ph/0511178.

\bibitem{Brodsky:1989qz}
  S.~J.~Brodsky and H.~J.~Lu,
  Phys.\ Rev.\ Lett.\  {\bf 64}, 1342 (1990).

\bibitem{Brodsky:2004qa}
  S.~J.~Brodsky, I.~Schmidt and J.~J.~Yang,
  Phys.\ Rev.\  D {\bf 70}, 116003 (2004)
  [arXiv:hep-ph/0409279].


\bibitem{Brodsky:1983vf}
  S.~J.~Brodsky, C.~R.~Ji and G.~P.~Lepage,
  Phys.\ Rev.\ Lett.\  {\bf 51}, 83 (1983).

\bibitem{Court:1986dh}
  G.~R.~Court {\it et al.},
  Phys.\ Rev.\ Lett.\  {\bf 57}, 507 (1986).

\bibitem{Brodsky:1987xw}
  S.~J.~Brodsky and G.~F.~de Teramond,
  Phys.\ Rev.\ Lett.\  {\bf 60}, 1924 (1988).

\bibitem{Brodsky:1982gc}
  S.~J.~Brodsky, G.~P.~Lepage and P.~B.~Mackenzie,
  Phys.\ Rev.\  D {\bf 28}, 228 (1983).
  
\bibitem{Grunberg:1991ac}
  G.~Grunberg and A.~L.~Kataev,
  Phys.\ Lett.\  B {\bf 279}, 352 (1992).

\bibitem{Brodsky:1998hn}
  S.~J.~Brodsky and D.~S.~Hwang,
  Nucl.\ Phys.\  B {\bf 543}, 239 (1999)
  [arXiv:hep-ph/9806358].

\bibitem{Brodsky:2000xy}
  S.~J.~Brodsky, M.~Diehl and D.~S.~Hwang,
  Nucl.\ Phys.\  B {\bf 596}, 99 (2001)
  [arXiv:hep-ph/0009254].


\bibitem{Brodsky:2006uq}
  S.~J.~Brodsky,
  Eur.\ Phys.\ J.\  A {\bf 31}, 638 (2007)
  [arXiv:hep-ph/0610115].

\bibitem{Grigoryan:2007my}
  H.~R.~Grigoryan and A.~V.~Radyushkin,
  arXiv:0706.1543 [hep-ph].

\bibitem{Brodsky:1994eh}
  S.~J.~Brodsky and H.~J.~Lu,
  Phys.\ Rev.\  D {\bf 51}, 3652 (1995)
  [arXiv:hep-ph/9405218].

\bibitem{Rathsman:1996jk}
  J.~Rathsman,
  Phys.\ Rev.\  D {\bf 54}, 3420 (1996)
  [arXiv:hep-ph/9605401].

\bibitem{Brodsky:1999gm}
  S.~J.~Brodsky and J.~Rathsman,
  ``Conformal symmetry as a template: Commensurate scale relations and
  physical renormalization schemes,''
  arXiv:hep-ph/9906339.

\bibitem{Brodsky:1995tb}
  S.~J.~Brodsky, G.~T.~Gabadadze, A.~L.~Kataev and H.~J.~Lu,
  Phys.\ Lett.\  B {\bf 372}, 133 (1996)
  [arXiv:hep-ph/9512367].

\bibitem{Brodsky:2000cr}
  S.~J.~Brodsky, E.~Gardi, G.~Grunberg and J.~Rathsman,
  Phys.\ Rev.\  D {\bf 63}, 094017 (2001)
  [arXiv:hep-ph/0002065].

\bibitem{Grunberg:1982fw}
  G.~Grunberg,
  Phys.\ Rev.\  D {\bf 29}, 2315 (1984).

\bibitem{Brodsky:1998mf}
  S.~J.~Brodsky, M.~S.~Gill, M.~Melles and J.~Rathsman,
  Phys.\ Rev.\  D {\bf 58}, 116006 (1998)
  [arXiv:hep-ph/9801330].

\bibitem{Binger:2003by}
  M.~Binger and S.~J.~Brodsky,
  Phys.\ Rev.\  D {\bf 69}, 095007 (2004)
  [arXiv:hep-ph/0310322].

\bibitem{Brodsky:1992pq}
  S.~J.~Brodsky and H.~J.~Lu,
  ``On the selfconsistency of scale setting methods,''
  arXiv:hep-ph/9211308.

\bibitem{Stueckelberg:1953dz}
  E.~C.~G.~Stueckelberg and A.~Petermann,
  Helv.\ Phys.\ Acta {\bf 26}, 499 (1953).

\bibitem{Brodsky:1997jk}
  S.~J.~Brodsky and P.~Huet,
  Phys.\ Lett.\  B {\bf 417}, 145 (1998)
  [arXiv:hep-ph/9707543].

\bibitem{Binger:2006sj}
  M.~Binger and S.~J.~Brodsky,
  Phys.\ Rev.\  D {\bf 74}, 054016 (2006)
  [arXiv:hep-ph/0602199].

\bibitem{Maldacena:1997re}
  J.~M.~Maldacena,
  Adv.\ Theor.\ Math.\ Phys.\  {\bf 2}, 231 (1998)
  [Int.\ J.\ Theor.\ Phys.\  {\bf 38}, 1113 (1999)]
  [arXiv:hep-th/9711200].

\bibitem{Polchinski:2001tt}
  J.~Polchinski and M.~J.~Strassler,
  Phys.\ Rev.\ Lett.\  {\bf 88}, 031601 (2002)
  [arXiv:hep-th/0109174].

\bibitem{Janik:1999zk}
  R.~A.~Janik and R.~Peschanski,
  Nucl.\ Phys.\  B {\bf 565}, 193 (2000)
  [arXiv:hep-th/9907177].

\bibitem{Erlich:2005qh}
  J.~Erlich, E.~Katz, D.~T.~Son and M.~A.~Stephanov,
  Phys.\ Rev.\ Lett.\  {\bf 95}, 261602 (2005)
  [arXiv:hep-ph/0501128].

\bibitem{Karch:2006pv}
  A.~Karch, E.~Katz, D.~T.~Son and M.~A.~Stephanov,
  Phys.\ Rev.\  D {\bf 74}, 015005 (2006)
  [arXiv:hep-ph/0602229].

\bibitem{deTeramond:2005su}
  G.~F.~de Teramond and S.~J.~Brodsky,
  Phys.\ Rev.\ Lett.\  {\bf 94}, 201601 (2005)
  [arXiv:hep-th/0501022].

\bibitem{Kovtun:2004de}
  P.~Kovtun, D.~T.~Son and A.~O.~Starinets,
  Phys.\ Rev.\ Lett.\  {\bf 94}, 111601 (2005)
  [arXiv:hep-th/0405231].

\bibitem{vonSmekal:1997is}
  L.~von Smekal, R.~Alkofer and A.~Hauck,
  Phys.\ Rev.\ Lett.\  {\bf 79}, 3591 (1997)
  [arXiv:hep-ph/9705242].

\bibitem{Zwanziger:2003cf}
  D.~Zwanziger,
  Phys.\ Rev.\  D {\bf 69}, 016002 (2004)
  [arXiv:hep-ph/0303028].

\bibitem{Mattingly:1993ej}
  A.~C.~Mattingly and P.~M.~Stevenson,
  Phys.\ Rev.\  D {\bf 49}, 437 (1994)
  [arXiv:hep-ph/9307266].

\bibitem{Brodsky:2002nb}
  S.~J.~Brodsky, S.~Menke, C.~Merino and J.~Rathsman,
  Phys.\ Rev.\  D {\bf 67}, 055008 (2003)
  [arXiv:hep-ph/0212078].

\bibitem{Baldicchi:2002qm}
  M.~Baldicchi and G.~M.~Prosperi,
  Phys.\ Rev.\  D {\bf 66}, 074008 (2002)
  [arXiv:hep-ph/0202172].

\bibitem{Brodsky:1998ua}
  S.~J.~Brodsky, J.~R.~Pelaez and N.~Toumbas,
  Phys.\ Rev.\  D {\bf 60}, 037501 (1999)
  [arXiv:hep-ph/9810424].

\bibitem{Furui:2006py}
  S.~Furui and H.~Nakajima,
  ``Infrared features of unquenched finite temperature lattice Landau gauge
  QCD,''
  arXiv:hep-lat/0612009.

\bibitem{Antonov:2007mx}
  D.~Antonov and H.~J.~Pirner,
  arXiv:hep-ph/0702227.

\bibitem{Brodsky:2007pt}
  S.~J.~Brodsky and G.~F.~de Teramond,
  arXiv:hep-th/0702205.

\bibitem{Parisi:1972zy}
  G.~Parisi,
  Phys.\ Lett.\  B {\bf 39}, 643 (1972).

\bibitem{Breitenlohner:1982jf}
  P.~Breitenlohner and D.~Z.~Freedman,
  Annals Phys.\  {\bf 144}, 249 (1982).

\bibitem{Brodsky:2005kc}
  S.~J.~Brodsky and G.~F.~de Teramond,
  AIP Conf.\ Proc.\  {\bf 814}, 108 (2006)
  [arXiv:hep-ph/0510240].

\bibitem{Burkardt:2005td}
  M.~Burkardt,
  Int.\ J.\ Mod.\ Phys.\  A {\bf 21}, 926 (2006)
  [arXiv:hep-ph/0509316].

\bibitem{Ji:2003ak}
  X.~d.~Ji,
  Phys.\ Rev.\ Lett.\  {\bf 91}, 062001 (2003)
  [arXiv:hep-ph/0304037].

\bibitem{Brodsky:2006in}
  S.~J.~Brodsky, D.~Chakrabarti, A.~Harindranath, A.~Mukherjee and J.~P.~Vary,
  Phys.\ Lett.\  B {\bf 641}, 440 (2006)
  [arXiv:hep-ph/0604262].

\bibitem{Hoyer:2006xg}
  P.~Hoyer,
  AIP Conf.\ Proc.\  {\bf 904}, 65 (2007)
  [arXiv:hep-ph/0608295].

\bibitem{Brodsky:2006ku}
  S.~J.~Brodsky, D.~Chakrabarti, A.~Harindranath, A.~Mukherjee and J.~P.~Vary,
  Phys.\ Rev.\  D {\bf 75}, 014003 (2007)
  [arXiv:hep-ph/0611159].

\bibitem{Lepage:1979zb}
  G.~P.~Lepage and S.~J.~Brodsky,
  Phys.\ Lett.\  B {\bf 87}, 359 (1979).

\bibitem{Choi:2006ha}
  H.~M.~Choi and C.~R.~Ji,
  Phys.\ Rev.\  D {\bf 74}, 093010 (2006)
  [arXiv:hep-ph/0608148].

\bibitem{Brodsky:1979qm}
  S.~J.~Brodsky and G.~P.~Lepage,
 ``Exclusive Processes And The Exclusive-Inclusive Connection In Quantum
 Chromodynamics,''  SLAC-PUB-2294.
Presented at the Workshop on Current Topics in High Energy Physics,
Cal Tech., Pasadena, Calif., Feb 13-17, 1979.

\bibitem{Brodsky:1974vy}
  S.~J.~Brodsky and G.~R.~Farrar,
  Phys.\ Rev.\  D {\bf 11}, 1309 (1975).

\bibitem{Matveev:1973ra}
  V.~A.~Matveev, R.~M.~Muradian and A.~N.~Tavkhelidze,
  Lett.\ Nuovo Cim.\  {\bf 7}, 719 (1973).

\bibitem{Brodsky:1989pv}
  S.~J.~Brodsky and G.~P.~Lepage,
  Adv.\ Ser.\ Direct.\ High Energy Phys.\  {\bf 5}, 93 (1989).

\bibitem{Holt:1990ze}
  R.~J.~Holt,
  Phys.\ Rev.\  C {\bf 41}, 2400 (1990).

\bibitem{Bochna:1998ca}
  C.~Bochna {\it et al.}  [E89-012 Collaboration],
  Phys.\ Rev.\ Lett.\  {\bf 81}, 4576 (1998)
  [arXiv:nucl-ex/9808001].

\bibitem{Rossi:2004qm}
  P.~Rossi {\it et al.}  [CLAS Collaboration],
  arXiv:hep-ph/0405207.

\bibitem{Danagoulian:2007gs}
  A.~Danagoulian {\it et al.}  [Hall A Collaboration],
  Phys.\ Rev.\ Lett.\  {\bf 98}, 152001 (2007)
  [arXiv:nucl-ex/0701068].


\bibitem{Chen:2001sm}
  A.~Chen,
  ``Proton Anti-Proton Pair Production In Two-Photon Collisions At Belle,''
Published in *Ascona 2001, The structure and interactions of the
photon* 227-230

\bibitem{Chen:2006ug}
  A.~E.~Chen  [BELLE Collaboration],
  Int.\ J.\ Mod.\ Phys.\  A {\bf 21}, 5543 (2006).

\bibitem{Berger:1981fr}
  E.~L.~Berger and S.~J.~Brodsky,
  Phys.\ Rev.\  D {\bf 24}, 2428 (1981).

\bibitem{Sivers:1975dg}
  D.~W.~Sivers, S.~J.~Brodsky and R.~Blankenbecler,
  Phys.\ Rept.\  {\bf 23}, 1 (1976).

\bibitem{Antreasyan:1978cw}
  D.~Antreasyan, J.~W.~Cronin, H.~J.~Frisch, M.~J.~Shochet, L.~Kluberg, P.~A.~Piroue and R.~L.~Sumner,
  Phys.\ Rev.\  D {\bf 19}, 764 (1979).

\bibitem{Tannenbaum:2006ku}
  M.~J.~Tannenbaum,
  ``Review of hard scattering and jet analysis,''
  arXiv:nucl-ex/0611008.

\bibitem{Adler:2003kg}
  S.~S.~Adler {\it et al.}  [PHENIX Collaboration],
  Phys.\ Rev.\ Lett.\  {\bf 91}, 172301 (2003)
  [arXiv:nucl-ex/0305036].

\bibitem{Franz:2000ee}
  M.~Franz, M.~V.~Polyakov and K.~Goeke,
  Phys.\ Rev.\  D {\bf 62}, 074024 (2000)
  [arXiv:hep-ph/0002240].

\bibitem{Brodsky:1984nx}
  S.~J.~Brodsky, J.~C.~Collins, S.~D.~Ellis, J.~F.~Gunion and A.~H.~Mueller,
Published in Snowmass Summer Study 1984:0227 (QCD184:S7:1984)

\bibitem{Harris:1995jx}
  B.~W.~Harris, J.~Smith and R.~Vogt,
  Nucl.\ Phys.\  B {\bf 461}, 181 (1996)
  [arXiv:hep-ph/9508403].

\bibitem{Brodsky:1997fj}
  S.~J.~Brodsky and M.~Karliner,
  Phys.\ Rev.\ Lett.\  {\bf 78}, 4682 (1997)
  [arXiv:hep-ph/9704379].

\bibitem{Brodsky:2001yt}
  S.~J.~Brodsky and S.~Gardner,
  Phys.\ Rev.\  D {\bf 65}, 054016 (2002)
  [arXiv:hep-ph/0108121].

\bibitem{Munger:1993kq}
  C.~T.~Munger, S.~J.~Brodsky and I.~Schmidt,
  Phys.\ Rev.\  D {\bf 49}, 3228 (1994).

\bibitem{Halzen:2004bn}
  F.~Halzen,
  Nucl.\ Phys.\ Proc.\ Suppl.\  {\bf 136}, 93 (2004)
  [Acta Phys.\ Polon.\  B {\bf 36}, 481 (2005\ APCPC,745,3-13.2005\ NUPHA,A752,3-13.2005)]
  [arXiv:astro-ph/0402083].

\bibitem{Dremin:2005dm}
  I.~M.~Dremin and V.~I.~Yakovlev,
  Astropart.\ Phys.\  {\bf 26}, 1 (2006)
  [arXiv:hep-ph/0510377].

\bibitem{Hoyer:1990us}
  P.~Hoyer, M.~Vanttinen and U.~Sukhatme,
  Phys.\ Lett.\  B {\bf 246}, 217 (1990).

\bibitem{Badier:1981ci}
  J.~Badier {\it et al.}  [NA3 Collaboration],
  Phys.\ Lett.\  B {\bf 104}, 335 (1981).

\bibitem{Leitch:1999ea}
  M.~J.~Leitch {\it et al.}  [FNAL E866/NuSea collaboration],
  Phys.\ Rev.\ Lett.\  {\bf 84}, 3256 (2000)
  [arXiv:nucl-ex/9909007].

\bibitem{Anjos:2001jr}
  J.~C.~Anjos, J.~Magnin and G.~Herrera,
  Phys.\ Lett.\  B {\bf 523}, 29 (2001)
  [arXiv:hep-ph/0109185].

\bibitem{Ocherashvili:2004hi}
  A.~Ocherashvili {\it et al.}  [SELEX Collaboration],
  Phys.\ Lett.\  B {\bf 628}, 18 (2005)
  [arXiv:hep-ex/0406033].

\bibitem{Vogt:1995tf}
  R.~Vogt and S.~J.~Brodsky,
  Phys.\ Lett.\  B {\bf 349}, 569 (1995)
  [arXiv:hep-ph/9503206].

\bibitem{Bari:1991ty}
  G.~Bari {\it et al.},
  Nuovo Cim.\  A {\bf 104}, 1787 (1991).

\bibitem{Brodsky:1976rz}
  S.~J.~Brodsky and B.~T.~Chertok,
  Phys.\ Rev.\  D {\bf 14}, 3003 (1976).

\bibitem{Matveev:1977xt}
  V.~A.~Matveev and P.~Sorba,
  Lett.\ Nuovo Cim.\  {\bf 20}, 435 (1977).

\bibitem{Arnold:1975dd}
  R.~G.~Arnold {\it et al.},
  Phys.\ Rev.\ Lett.\  {\bf 35}, 776 (1975).

\bibitem{Farrar:1991qi}
  G.~R.~Farrar, K.~Huleihel and H.~y.~Zhang,
  Phys.\ Rev.\ Lett.\  {\bf 74}, 650 (1995).

\bibitem{Asner:2006pw}
  D.~M.~Asner {\it et al.}  [CLEO Collaboration],
  Phys.\ Rev.\  D {\bf 75}, 012009 (2007)
  [arXiv:hep-ex/0612019].

\bibitem{Adloff:1997sc}
  C.~Adloff {\it et al.}  [H1 Collaboration],
  Z.\ Phys.\  C {\bf 76}, 613 (1997)
  [arXiv:hep-ex/9708016].

\bibitem{Breitweg:1998gc}
  J.~Breitweg {\it et al.}  [ZEUS Collaboration],
  Eur.\ Phys.\ J.\  C {\bf 6}, 43 (1999)
  [arXiv:hep-ex/9807010].

\bibitem{Brodsky:1994kf}
  S.~J.~Brodsky, L.~Frankfurt, J.~F.~Gunion, A.~H.~Mueller and M.~Strikman,
  Phys.\ Rev.\  D {\bf 50}, 3134 (1994)
  [arXiv:hep-ph/9402283].

\bibitem{Belitsky:2002sm}
  A.~V.~Belitsky, X.~Ji and F.~Yuan,
  Nucl.\ Phys.\  B {\bf 656}, 165 (2003)
  [arXiv:hep-ph/0208038].

\bibitem{Collins:2004nx}
  J.~C.~Collins and A.~Metz,
  Phys.\ Rev.\ Lett.\  {\bf 93}, 252001 (2004)
  [arXiv:hep-ph/0408249].

\bibitem{Collins:2003fm}
  J.~C.~Collins,
  Acta Phys.\ Polon.\  B {\bf 34}, 3103 (2003)
  [arXiv:hep-ph/0304122].

\bibitem{Brodsky:2004hi}
  S.~J.~Brodsky, R.~Enberg, P.~Hoyer and G.~Ingelman,
  Phys.\ Rev.\  D {\bf 71}, 074020 (2005)
  [arXiv:hep-ph/0409119].

\bibitem{Edin:1995gi}
  A.~Edin, G.~Ingelman and J.~Rathsman,
  Phys.\ Lett.\  B {\bf 366}, 371 (1996)
  [arXiv:hep-ph/9508386].

\bibitem{Burkardt:2004vm}
  M.~Burkardt,
  Nucl.\ Phys.\ Proc.\ Suppl.\  {\bf 141}, 86 (2005)
  [arXiv:hep-ph/0408009].

\bibitem{Avakian:2004qt}
  H.~Avakian and L.~Elouadrhiri  [CLAS Collaboration],
  AIP Conf.\ Proc.\  {\bf 698}, 612 (2004).

\bibitem{Brodsky:2002rv}
  S.~J.~Brodsky, D.~S.~Hwang and I.~Schmidt,
  Nucl.\ Phys.\  B {\bf 642}, 344 (2002)
  [arXiv:hep-ph/0206259].

\bibitem{Brodsky:2002pr}
  S.~J.~Brodsky, D.~S.~Hwang and I.~Schmidt,
  Phys.\ Lett.\  B {\bf 553}, 223 (2003)
  [arXiv:hep-ph/0211212].

\bibitem{Brodsky:2006ha}
  S.~J.~Brodsky and S.~Gardner,
  Phys.\ Lett.\  B {\bf 643}, 22 (2006)
  [arXiv:hep-ph/0608219].

\bibitem{Ji:2002aa}
  X.~d.~Ji and F.~Yuan,
  Phys.\ Lett.\  B {\bf 543}, 66 (2002)
  [arXiv:hep-ph/0206057].

\bibitem{Bertsch:1981py}
  G.~Bertsch, S.~J.~Brodsky, A.~S.~Goldhaber and J.~F.~Gunion,
  Phys.\ Rev.\ Lett.\  {\bf 47}, 297 (1981).

\bibitem{Frankfurt:1999tq}
  L.~Frankfurt, G.~A.~Miller and M.~Strikman,
  Found.\ Phys.\  {\bf 30}, 533 (2000)
  [arXiv:hep-ph/9907214].

\bibitem{Nikolaev:2000sh}
  N.~N.~Nikolaev, W.~Schafer and G.~Schwiete,
  Phys.\ Rev.\  D {\bf 63}, 014020 (2001)
  [arXiv:hep-ph/0009038].

\bibitem{Aitala:2000hc}
  E.~M.~Aitala {\it et al.}  [E791 Collaboration],
  Phys.\ Rev.\ Lett.\  {\bf 86}, 4773 (2001)
  [arXiv:hep-ex/0010044].

\bibitem{Borisov:2002rd}
  A.~B.~Borisov  [HERMES Collaboration],
  Nucl.\ Phys.\  A {\bf 711}, 269 (2002).

\bibitem{Aclander:2004zm}
  J.~L.~S.~Aclander {\it et al.},
  Phys.\ Rev.\  C {\bf 70}, 015208 (2004)
  [arXiv:nucl-ex/0405025].

\bibitem{Ioffe:1969kf}
  B.~L.~Ioffe,
  Phys.\ Lett.\  B {\bf 30}, 123 (1969).

\bibitem{Zeller:2001hh}
  G.~P.~Zeller {\it et al.}  [NuTeV Collaboration],
  Phys.\ Rev.\ Lett.\  {\bf 88}, 091802 (2002)
  [Erratum-ibid.\  {\bf 90}, 239902 (2003)]
  [arXiv:hep-ex/0110059].

\bibitem{Brodsky:2000sk}
  S.~J.~Brodsky,
  arXiv:hep-ph/0004211.
  Presented at the Workshop on Light-Cone QCD and NonPerturbative Hadron Physics, Adelaide, Australia, 13-22 Dec 1999.
Published in Adelaide 1999, Lightcone QCD and nonperturbative hadron
physics* 15-26.


\bibitem{Brodsky:1996hc}
  S.~J.~Brodsky and B.~Q.~Ma,
  Phys.\ Lett.\  B {\bf 381}, 317 (1996)
  [arXiv:hep-ph/9604393].

\bibitem{Kretzer:2004bg}
  S.~Kretzer,
  arXiv:hep-ph/0408287.

\bibitem{Portheault:2004xy}
  B.~Portheault,
  arXiv:hep-ph/0406226.

\bibitem{Olness:2003wz}
  F.~Olness {\it et al.},
  Eur.\ Phys.\ J.\  C {\bf 40}, 145 (2005)
  [arXiv:hep-ph/0312323].

\bibitem{Rathsman:2001xe}
  J.~Rathsman,
  ``Conformal expansions: A template for QCD predictions,''
in {\it Proc. of the 5th International Symposium on Radiative Corrections (RADCOR 2000) } ed. Howard E. Haber,
  arXiv:hep-ph/0101248.

\bibitem{Grunberg:2001bz}
  G.~Grunberg,
  JHEP {\bf 0108}, 019 (2001)
  [arXiv:hep-ph/0104098].

\bibitem{Banks:1981nn}
  T.~Banks and A.~Zaks,
  Nucl.\ Phys.\  B {\bf 196}, 189 (1982).

\bibitem{Lepage:1980fj}
  G.~P.~Lepage and S.~J.~Brodsky,
  Phys.\ Rev.\  D {\bf 22}, 2157 (1980).

\bibitem{Crewther:1972kn}
  R.~J.~Crewther,
  Phys.\ Rev.\ Lett.\  {\bf 28}, 1421 (1972).

\bibitem{Brodsky:1980ny}
  S.~J.~Brodsky, Y.~Frishman, G.~P.~Lepage and C.~T.~Sachrajda,
  Phys.\ Lett.\  B {\bf 91}, 239 (1980).

\bibitem{Brodsky:1984xk}
  S.~J.~Brodsky, P.~Damgaard, Y.~Frishman and G.~P.~Lepage,
  Phys.\ Rev.\  D {\bf 33}, 1881 (1986).

\bibitem{Brodsky:1985ve}
  S.~J.~Brodsky, Y.~Frishman and G.~P.~Lepage,
  Phys.\ Lett.\  B {\bf 167}, 347 (1986).

\bibitem{Braun:2003rp}
  V.~M.~Braun, G.~P.~Korchemsky and D.~Mueller,
  Prog.\ Part.\ Nucl.\ Phys.\  {\bf 51}, 311 (2003)
  [arXiv:hep-ph/0306057].

\bibitem{Brodsky:2004qb}
  S.~J.~Brodsky,
  ``Conformal symmetry as a template for QCD,''
  arXiv:hep-ph/0408069.

\bibitem{Brodsky:2003dn}
  S.~J.~Brodsky,
  ``Conformal aspects of QCD,''
SLAC-PUB-10206.  Published in *Wako 2003, Color confinement and
hadrons in quantum chromodynamics* 164-175


\bibitem{Brodsky:2000dr}
  S.~J.~Brodsky,
 ``Exclusive processes in quantum chromodynamics and the light-cone Fock
  representation,''
  SLAC-PUB-8649.
  Contribution to the Boris Ioffe Festschrift 'At the Frontier of Particle Physics: A Handbook for QCD,' edited by M. Shifman.
In *Shifman, M. (ed.): At the frontier of particle physics, vol. 2*
1343-1444.

\bibitem{Andersson:1983ia}
  B.~Andersson, G.~Gustafson, G.~Ingelman and T.~Sjostrand,
  Phys.\ Rept.\  {\bf 97}, 31 (1983).

\bibitem{Brodsky:2007}
  S.~J.~Brodsky and G.~F.~de Teramond to be published.




\end{thebibliography}
\end{document}